\documentstyle[12pt,epsfig]{article}
\def\thefootnote{\fnsymbol{footnote}}

\begin{document}

\vspace{0.2cm}

\begin{center}
{\large \bf Two-zero Textures of the Majorana Neutrino Mass Matrix
and Current Experimental Tests}
\end{center}

\vspace{0.2cm}
\begin{center}
{\bf Harald Fritzsch} $^a$, ~
{\bf Zhi-zhong Xing} $^b$ \footnote{E-mail: xingzz@ihep.ac.cn}, ~
{\bf Shun Zhou} $^c$ \footnote{E-mail: zhoush@mppmu.mpg.de} \\
$^a${\sl Physik-Department, Ludwig-Maximilians-Universit\"{a}t,
80333 M\"{u}nchen, Germany} \\
$^b${\sl Institute of High Energy Physics, Chinese Academy of
Sciences, Beijing 100049, China} \\
$^c${\sl Max-Planck-Institut f\"{u}r Physik
(Werner-Heisenberg-Institut), 80805 M\"{u}nchen, Germany}
\end{center}

\vspace{2.0cm}

\begin{abstract}
In view of the latest T2K and MINOS neutrino oscillation data which
hint at a relatively large $\theta^{}_{13}$, we perform a systematic
study of the Majorana neutrino mass matrix $M^{}_\nu$ with two
independent texture zeros. We show that three neutrino masses
$(m^{}_1, m^{}_2, m^{}_3)$ and three CP-violating phases $(\delta,
\rho, \sigma)$ can fully be determined from two neutrino
mass-squared differences $(\delta m^2, \Delta m^2)$ and three flavor
mixing angles $(\theta^{}_{12}, \theta^{}_{23}, \theta^{}_{13})$. We
find that seven patterns of $M^{}_\nu$ (i.e., $\bf A^{}_{1,2}$, $\bf
B^{}_{1,2,3,4}$ and ${\bf C}$) are compatible with current
experimental data at the $3\sigma$ level, but the parameter space of
each pattern is more strictly constrained than before. We
demonstrate that the texture zeros of $M^{}_\nu$ are stable against
the one-loop quantum corrections, and there exists a permutation
symmetry between Patterns $\bf A^{}_1$ and $\bf A^{}_2$, $\bf
B^{}_1$ and $\bf B^{}_2$ or $\bf B^{}_3$ and $\bf B^{}_4$.
Phenomenological implications of $M^{}_\nu$ on the neutrinoless
double-beta decay and leptonic CP violation are discussed, and a
realization of those texture zeros by means of the $Z^{}_n$ flavor
symmetries is illustrated.
\end{abstract}
\begin{center}
\end{center}

\newpage

\setcounter{footnote}{0}
\def\thefootnote{\arabic{footnote}}

\section{Introduction}

Compelling evidence in favor of neutrino oscillations has been
accumulated from a number of solar, atmospheric, reactor and
accelerator neutrino experiments since 1998 \cite{PDG}. We are now
convinced that the three known neutrinos have nonzero and
nondegenerate masses, and their flavor states can convert from one
kind to another. In particular, two neutrino mass-squared
differences $(\delta m^2, \Delta m^2)$ and two flavor mixing angles
$(\theta^{}_{12}, \theta^{}_{23})$ have been determined to a
reasonably good degree of accuracy from currently available
experimental data. In spite of this progress made in neutrino
physics, our quantitative knowledge about the properties of massive
neutrinos remain quite incomplete --- for instance, the absolute
mass scale of three neutrinos, the sign of $\Delta m^2$ and the
smallest flavor mixing angle $\theta^{}_{13}$ are still unknown.
Although the nature of massive neutrinos is also an open question,
we assume them to be the Majorana particles. In this case the
$3\times 3$ Majorana neutrino mass matrix $M^{}_\nu$ is symmetric,
and the corresponding $3\times 3$ flavor mixing matrix $V$ contains
three CP-violating phases $(\delta, \rho, \sigma)$ which are
entirely unrestricted by current experimental data.

The T2K \cite{T2K} and MINOS \cite{MINOS} accelerator neutrino
experiments have recently announced their preliminary data on the
appearance $\nu^{}_\mu \to \nu^{}_e$ oscillation, which hints at a
relatively large value of $\theta^{}_{13}$. If this mixing angle is
confirmed to be not very small by new data from these two
experiments and from the upcoming reactor antineutrino experiments
(e.g., the Double Chooz experiment in France \cite{DC}, the Daya Bay
experiment in China \cite{DYB} and the RENO experiment in Korea
\cite{RENO}), it will be a great news to the long-baseline neutrino
oscillation experiments which aim at a determination of the sign of
$\Delta m^2$ and the magnitude of the Dirac CP-violating phase
$\delta$. The observed values of relevant neutrino flavor parameters
may then allow us to reconstruct the Majorana neutrino mass matrix
$M^{}_\nu$ in the flavor basis where the charged-lepton mass matrix
$M^{}_l$ is diagonal: $M^{}_\nu = V \widehat{M}^{}_\nu V^T$, where
$\widehat{M}^{}_\nu = {\rm Diag}\{m^{}_1, m^{}_2, m^{}_3\}$ with
$m^{}_i$ (for $i=1,2,3$) being the neutrino masses, and $V$ is just
the $3\times 3$ neutrino mixing matrix.

In the lack of a convincing flavor theory, at least five approaches
have so far been tried to study the flavor problems of massive
neutrinos \cite{X04}: (a) radiative mechanisms, (b) texture zeros,
(c) flavor symmetries, (d) seesaw mechanisms, and (e) extra
dimensions. Some of them are certainly correlative. For example, the
neutrino mass matrix $M^{}_\nu$ may have a few texture zeros as a
natural consequence of an underlying flavor symmetry in a given
model with or without the seesaw mechanism. Such texture zeros are
phenomenologically useful in the sense that they guarantee the
calculability of $M^{}_\nu$ from which both the neutrino mass
spectrum and the flavor mixing pattern can more or less be
predicted. Note that texture zeros of a fermion mass matrix
dynamically mean that the corresponding matrix elements are
sufficiently suppressed in comparison with their neighboring
counterparts \cite{FN}, and they can help us to establish some
simple and testable relations between flavor mixing angles and
fermion mass ratios --- if such relations turn out to be favored by
the experimental data, they might have a fundamental reason and
should originate from the underlying flavor theory. Hence a
phenomenological study of possible texture zeros of $M^{}_\nu$ {\it
does} make a lot of sense.

As $M^{}_\nu$ is symmetric, it has six independent complex entries.
If $n$ of them are taken to be vanishing (i.e., $M^{}_\nu$ has $n$
independent texture zeros), then we shall arrive at
\begin{eqnarray}
^6{\bf C}_n = \frac{6!}{n! \left (6 - n\right )!}
\end{eqnarray}
different textures. It is easy to show that a texture of $M^{}_\nu$
with more than two independent zeros (i.e., $n\geq 3$) is definitely
incompatible with current experimental data on neutrino masses and
flavor mixing angles \cite{Xing04}. Hence a number of authors have
paid particular interest to the two-zero textures of $M^{}_\nu$
\cite{FGM}---\cite{More}
\footnote{It is worth mentioning that the one-zero textures of
$M^{}_\nu$ have much less predictability than the two-zero textures
of $M^{}_\nu$, and their phenomenological implications have been
discussed in Ref. \cite{Xing05}.}.
There are totally fifteen two-zero textures of $M^{}_\nu$, which can
be classified into six categories:
\begin{eqnarray}
{\bf A^{}_1}: ~~ \left(\matrix{ 0 & 0 & \times \cr 0 & \times &
\times \cr \times & \times & \times \cr} \right) \; , ~~~~ {\bf
A^{}_2}: ~~ \left(\matrix{ 0 & \times & 0 \cr \times & \times &
\times \cr 0 & \times & \times \cr} \right) \; ;
\end{eqnarray}
\begin{eqnarray}
{\bf B^{}_1}: ~~ \left(\matrix{ \times & \times & 0 \cr \times & 0 &
\times \cr 0 & \times & \times \cr} \right) \; , && {\bf B^{}_2}: ~~
\left(\matrix{ \times & 0 & \times \cr 0 & \times & \times \cr
\times & \times & 0 \cr} \right) \; , ~~~~~
\nonumber \\
{\bf B^{}_3}: ~~ \left(\matrix{ \times & 0 & \times \cr 0 & 0 &
\times \cr \times & \times & \times \cr} \right) \; , && {\bf
B^{}_4}: ~~ \left(\matrix{ \times & \times & 0 \cr \times & \times &
\times \cr 0 & \times & 0 \cr} \right) \; ;
\end{eqnarray}
\begin{eqnarray}
{\bf C}: ~~ \left(\matrix{ \times & \times & \times \cr \times & 0 &
\times \cr \times & \times & 0 \cr} \right) \; ;
\end{eqnarray}
\begin{eqnarray}
{\bf D^{}_1}: ~~ \left(\matrix{ \times & \times & \times \cr \times
& 0 & 0 \cr \times & 0 & \times \cr} \right) \; , ~~~~ {\bf D^{}_2}:
~~ \left(\matrix{ \times & \times & \times \cr \times & \times & 0
\cr \times & 0 & 0 \cr} \right) \; ;
\end{eqnarray}
\begin{eqnarray}
{\bf E^{}_1}: ~~ \left(\matrix{ 0 & \times & \times \cr \times & 0 &
\times \cr \times & \times & \times \cr} \right) \; , ~~~~ {\bf
E^{}_2}: ~~ \left(\matrix{ 0 & \times & \times \cr \times & \times &
\times \cr \times & \times & 0 \cr} \right) \; , ~~~~ {\bf E^{}_3}:
~~ \left(\matrix{ 0 & \times & \times \cr \times & \times & 0 \cr
\times & 0 & \times \cr} \right) \; ;
\end{eqnarray}
and
\begin{eqnarray}
{\bf F^{}_1}: ~~ \left(\matrix{ \times & 0 & 0 \cr 0 & \times &
\times \cr 0 & \times & \times \cr} \right) \; , ~~~~ {\bf F^{}_2}:
~~ \left(\matrix{ \times & 0 & \times \cr 0 & \times & 0 \cr \times
& 0 & \times \cr} \right) \; , ~~~~ {\bf F^{}_3}: ~~ \left(\matrix{
\times & \times & 0 \cr \times & \times & 0 \cr 0 & 0 & \times \cr}
\right) \; ,
\end{eqnarray}
in which each ``$\times$" denotes a nonzero matrix element. Previous
analyses of these fifteen patterns (see, e.g., Ref. \cite{Guo}) led
us to the following conclusions: (1) seven patterns (i.e., $\bf
A^{}_{1,2}$, $\bf B^{}_{1,2,3,4}$ and $\bf C$) were
phenomenologically favored; (2) two patterns (i.e., $\bf
D^{}_{1,2}$) were only marginally allowed; and (3) six patterns
(i.e., $\bf E^{}_{1,2,3}$ and $\bf F^{}_{1,2,3}$) were ruled out.
Taking account of the latest T2K and MINOS neutrino oscillation
data, one may wonder whether the above conclusions remain true or
not. A fast check tells us that Patterns $\bf D^{}_{1,2}$ can be
ruled out by today's experimental data at the $3\sigma$ level,
simply because $\theta^{}_{12} > 38^\circ$ is no more favored. In
addition, Patterns $\bf E^{}_{1,2,3}$ and $\bf F^{}_{1,2,3}$ remain
phenomenologically disfavored. The question turns out to be whether
Patterns $\bf A^{}_{1,2}$, $\bf B^{}_{1,2,3,4}$ and $\bf C$ can
survive current experimental tests and coincide with a relatively
large value of $\theta^{}_{13}$. The main purpose of this paper is
just to answer this question.

We aim to perform a very systematic study of the aforementioned
seven patterns of $M^{}_\nu$ with two independent texture zeros
based on a global fit of current neutrino oscillation data done by
Fogli {\it et al} \cite{Fogli}
\footnote{Note that Schwetz {\it et al} have done another global fit
of current data \cite{Schwetz}. Although their best-fit results of
three flavor mixing angles are slightly different from those
obtained by Fogli {\it et al} \cite{Fogli}, such differences become
insignificant at the $3\sigma$ level. Therefore, we shall mainly use
the $3\sigma$ results of \cite{Fogli} in our numerical
calculations.}.
Because a two-zero texture of $M^{}_\nu$ is very simple and can
reveal the salient phenomenological features of a given Majorana
neutrino mass matrix, we intend to take such a systematic analysis
as a good example to show how to test possible textures of
$M^{}_\nu$ by confronting them with more and more accurate
experimental data. Hence this work is different from any of the
previous ones. In particular, almost all the physical consequences
of Patterns $\bf A^{}_{1,2}$, $\bf B^{}_{1,2,3,4}$ and $\bf C$ are
explored in an analytical way in the present work; the stability of
texture zeros and the permutation symmetry between two similar
patterns are discussed; the numerical analysis is most updated and
more complete; and a realization of texture zeros by mean of
certain flavor symmetries is illustrated. We find that Patterns $\bf
A^{}_{1,2}$, $\bf B^{}_{1,2,3,4}$ and $\bf C$ of $M^{}_\nu$ can all
survive current experimental tests at the $3\sigma$ level, but we
believe that some of them are likely to be ruled out in the near future.

The remaining parts of this paper are organized as follows. In
section 2 we show that the full neutrino mass spectrum and two
Majorana CP-violating phases $(\rho, \sigma)$ can all be determined
in terms of three flavor mixing angles $(\theta^{}_{12},
\theta^{}_{23}, \theta^{}_{13})$ and the Dirac CP-violating phase
($\delta$) for a given two-zero texture of $M^{}_\nu$. Section 3 is
devoted to a complete analytical analysis of Patterns $\bf
A^{}_{1,2}$, $\bf B^{}_{1,2,3,4}$ and $\bf C$ of $M^{}_\nu$, and to
some discussions about the stability of texture zeros against the
one-loop quantum corrections and the permutation symmetry between
two similar patterns under consideration. With the help of current
neutrino oscillation data we perform a detailed numerical analysis
of those typical patterns of $M^{}_\nu$ in section 4. Section 5 is
devoted to some discussions about how to realize texture zeros of
$M^{}_\nu$ by means of certain flavor symmetries, and section 6 is a
summary of this work together with some concluding remarks.

\section{Basic Formulas}

\subsection{Important relations}

In the flavor basis where the charged-lepton mass matrix $M^{}_l$ is
diagonal, the Majorana neutrino mass matrix $M^{}_\nu$ can be
reconstructed in terms of three neutrino masses $(m^{}_1, m^{}_2,
m^{}_3)$ and the flavor mixing matrix $V$. Namely,
\begin{equation}
M^{}_\nu = V \left(\matrix{ m^{}_1 & 0 & 0 \cr 0 & m^{}_2 & 0 \cr 0
& 0 & m^{}_3}\right) V^{\rm T} \; .
\end{equation}
It is convenient to express $V$ as $V = U P$, where $U$ denotes a
$3\times 3$ unitary matrix consisting of three flavor mixing angles
($\theta^{}_{12}, \theta^{}_{23}, \theta^{}_{13}$) and one Dirac
CP-violating phase ($\delta$), and $P = {\rm Diag}\{e^{i\rho},
e^{i\sigma}, 1\}$ is a diagonal phase matrix containing two Majorana
CP-violating phases ($\rho, \sigma$). More explicitly, we adopt the
parametrization
\begin{equation}
U = \left(\matrix{c^{}_{12} c^{}_{13} & s^{}_{12} c^{}_{13} &
s^{}_{13} \cr -c^{}_{12} s^{}_{23} s^{}_{13} - s^{}_{12} c^{}_{23}
e^{-i\delta} & -s^{}_{12} s^{}_{23} s^{}_{13} + c^{}_{12} c^{}_{23}
e^{-i\delta} & s^{}_{23} c^{}_{13} \cr -c^{}_{12} c^{}_{23}
s^{}_{13} + s^{}_{12} s^{}_{23} e^{-i\delta} & -s^{}_{12} c^{}_{23}
s^{}_{13} - c^{}_{12} s^{}_{23} e^{-i\delta} & c^{}_{23}
c^{}_{13}}\right)  \; ,
\end{equation}
where $s^{}_{ij} \equiv \sin \theta^{}_{ij}$ and $c^{}_{ij} \equiv
\cos \theta^{}_{ij}$ (for $ij = 12, 23, 13$) are defined. The
neutrino mass matrix $M^{}_\nu$ can equivalently be written as
\begin{equation}
M^{}_\nu = U \left(\matrix{ \lambda^{}_1 & 0 & 0 \cr 0 &
\lambda^{}_2 & 0 \cr 0 & 0 & \lambda^{}_3}\right) U^{\rm T} \; ,
\end{equation}
where $\lambda^{}_1 = m^{}_1 e^{2i\rho}$, $\lambda^{}_2 = m^{}_2
e^{2i\sigma}$ and $\lambda^{}_3 = m^{}_3$. If two independent
elements of $M^{}_\nu$ are vanishing (i.e., $(M^{}_\nu)^{}_{ab} =
(M^{}_\nu)^{}_{\alpha \beta} = 0$ with $ab \neq \alpha \beta$) as
shown in Eqs. (2)---(7), one can obtain \cite{xing1}
\begin{eqnarray}
\frac{\lambda^{}_1}{\lambda^{}_3} &=&  \frac{U^{}_{a3} U^{}_{b3}
U^{}_{\alpha 2} U^{}_{\beta 2} - U^{}_{a2} U^{}_{b2} U^{}_{\alpha 3}
U^{}_{\beta 3}}{U^{}_{a2} U^{}_{b2} U^{}_{\alpha 1} U^{}_{\beta 1} -
U^{}_{a1} U^{}_{b1} U^{}_{\alpha 2} U^{}_{\beta 2}} \; ,
\nonumber \\
\frac{\lambda^{}_2}{\lambda^{}_3} &=& \frac{U^{}_{a1} U^{}_{b1}
U^{}_{\alpha 3} U^{}_{\beta 3} - U^{}_{a3} U^{}_{b3} U^{}_{\alpha 1}
U^{}_{\beta 1}}{U^{}_{a2} U^{}_{b2} U^{}_{\alpha 1} U^{}_{\beta 1} -
U^{}_{a1} U^{}_{b1} U^{}_{\alpha 2} U^{}_{\beta 2}} \; ,
\end{eqnarray}
from which two neutrino mass ratios are given by
\begin{eqnarray}
\xi &\equiv& \frac{m^{}_1}{m^{}_3} = \left|\frac{U^{}_{a3} U^{}_{b3}
U^{}_{\alpha 2} U^{}_{\beta 2} - U^{}_{a2} U^{}_{b2} U^{}_{\alpha 3}
U^{}_{\beta 3}}{U^{}_{a2} U^{}_{b2} U^{}_{\alpha 1} U^{}_{\beta 1} -
U^{}_{a1} U^{}_{b1} U^{}_{\alpha 2} U^{}_{\beta 2}}\right| \; ,
\nonumber \\
\zeta &\equiv& \frac{m^{}_2}{m^{}_3} = \left|\frac{U^{}_{a1}
U^{}_{b1} U^{}_{\alpha 3} U^{}_{\beta 3} - U^{}_{a3} U^{}_{b3}
U^{}_{\alpha 1} U^{}_{\beta 1}}{U^{}_{a2} U^{}_{b2} U^{}_{\alpha 1}
U^{}_{\beta 1} - U^{}_{a1} U^{}_{b1} U^{}_{\alpha 2} U^{}_{\beta
2}}\right| \; ,
\end{eqnarray}
and two Majorana CP-violating phases turn out to be
\begin{eqnarray}
\rho &=& \frac{1}{2} \arg \left[\frac{U^{}_{a3} U^{}_{b3}
U^{}_{\alpha 2} U^{}_{\beta 2} - U^{}_{a2} U^{}_{b2} U^{}_{\alpha 3}
U^{}_{\beta 3}}{U^{}_{a2} U^{}_{b2} U^{}_{\alpha 1} U^{}_{\beta 1} -
U^{}_{a1} U^{}_{b1} U^{}_{\alpha 2} U^{}_{\beta 2}}\right] \; ,
\nonumber \\
\sigma &=& \frac{1}{2} \arg \left[\frac{U^{}_{a1} U^{}_{b1}
U^{}_{\alpha 3} U^{}_{\beta 3} - U^{}_{a3} U^{}_{b3} U^{}_{\alpha 1}
U^{}_{\beta 1}}{U^{}_{a2} U^{}_{b2} U^{}_{\alpha 1} U^{}_{\beta 1} -
U^{}_{a1} U^{}_{b1} U^{}_{\alpha 2} U^{}_{\beta 2}}\right] \; .
\end{eqnarray}

\subsection{Parameter counting}

Since we have assumed massive neutrinos to be the Majorana
particles, there are nine physical parameters: three neutrino masses
$(m^{}_1, m^{}_2, m^{}_3)$, three flavor mixing angles
$(\theta^{}_{12}, \theta^{}_{23}, \theta^{}_{13})$, and three
CP-violating phases $(\delta, \rho, \sigma)$. By imposing two
independent texture zeros on $M^{}_\nu$, we obtain four constraint
relations as given in Eqs. (12) and (13). With the help of current
experimental data on three flavor mixing angles and two independent
neutrino mass-squared differences, defined as \cite{Fogli}
\begin{eqnarray}
\delta m^2 \equiv m^2_2 - m^2_1 \; , ~~~~ \Delta m^2 = m^2_3 -
\frac{1}{2} \left(m^2_1 + m^2_2\right) \; ,
\end{eqnarray}
one may determine or constrain both the neutrino mass spectrum and
three CP-violating phases through Eqs. (12) and (13).

We take $(\theta^{}_{12}, \theta^{}_{23}, \theta^{}_{13})$ and
$(\delta m^2, \Delta m^2)$ as observables to see why the neutrino
mass spectrum and three CP-violating phases can in principle be
determined or constrained. First of all, note that $\xi =
m^{}_1/m^{}_3$ and $\zeta = m^{}_2/m^{}_3$ are functions of the
Dirac CP-violating phase $\delta$ as shown in Eq. (12). Then it is
possible to determine or constrain $\delta$ from the relation
\begin{equation}
R^{}_\nu \equiv \frac{\delta m^2}{\left|\Delta m^2\right|} = \frac{2
\left(\zeta^2 - \xi^2\right)}{\left|2 - \left(\zeta^2 +
\xi^2\right)\right|} \; .
\end{equation}
Once $\delta$ is fixed, we can obtain $(\xi, \zeta)$ and $(\rho,
\sigma)$ from Eq. (12) and Eq. (13), respectively. In addition we
have
\begin{eqnarray}
m^{}_3 = \frac{\sqrt{\delta m^2}}{\sqrt{\zeta^2 - \xi^2}} \; , ~~~~
m^{}_2 = m^{}_3 \zeta \; , ~~~~ m^{}_1 = m^{}_3 \xi \; .
\end{eqnarray}
Thus the neutrino mass spectrum is fully determined.

Table 1 shows the global-fit values of $(\theta^{}_{12},
\theta^{}_{23}, \theta^{}_{13})$ and $(\delta m^2, \Delta m^2)$
obtained by assuming CP conservation with $\cos \delta = \pm 1$
\cite{Fogli}. We allow these parameters to independently vary in
their $3\sigma$ ranges:
\begin{eqnarray}
0.259 \leq \sin^2 \theta^{}_{12} \leq 0.359 \;  ~~&{\rm or}&~~
30.6^\circ \leq \theta^{}_{12} \leq 36.8^\circ \; ,
\nonumber \\
0.340 \leq \sin^2 \theta^{}_{23} \leq 0.640 \;  ~~&{\rm or}&~~
35.7^\circ \leq \theta^{}_{23} \leq 53.1^\circ \; ,
\nonumber \\
0.001 \leq \sin^2 \theta^{}_{13} \leq 0.044 \; ~~&{\rm or}&~~
~~1.8^\circ \leq \theta^{}_{13} \leq 12.1^\circ \; ;
\end{eqnarray}
and
\begin{eqnarray}
6.99\times 10^{-5}~{\rm eV}^2 \leq &\delta m^2& \leq 8.18 \times
10^{-5}~{\rm eV}^2 \; ,
\nonumber \\
2.06\times 10^{-3}~{\rm eV}^2 \leq & \hspace{-0.25cm}
|\Delta m^2| \hspace{-0.25cm} & \leq 2.67 \times
10^{-3}~{\rm eV}^2 \; .
\end{eqnarray}
In our numerical calculations we shall treat $\delta$ as an
unconstrained parameter. Note that the sign of $\Delta m^2$ remains
unknown: $\Delta m^2 > 0$ or $\Delta m^2 < 0$ corresponds to the
normal or inverted mass hierarchy of three neutrinos.
\begin{table}[t]
\caption{The latest global-fit results of three neutrino mixing
angles $(\theta^{}_{12}, \theta^{}_{23}, \theta^{}_{13})$ and two
neutrino mass-squared differences $\delta m^2$ and $\Delta m^2$
defined in Eq. (14). Here $\cos \delta = \pm 1$, and the old reactor
antineutrino fluxes have been assumed \cite{Fogli}.}
\begin{center}
\begin{tabular}{cccccc}
  \hline
  \hline
  Parameter & $\delta m^2~(10^{-5}~{\rm eV}^2)$ & $\Delta m^2~(10^{-3}~{\rm eV}^2)$
  & $\theta^{}_{12}$ & $\theta^{}_{23}$ & $\theta^{}_{13}$ \\
  \hline
  Best fit & $7.58$ & $2.35$ & $33.6^\circ$ & $40.4^\circ$ & $8.3^\circ$ \\
  $1\sigma$ range & $[7.32, 7.80]$ & $[2.26, 2.47]$ & $[32.6^\circ, 34.7^\circ]$
  & $[38.6^\circ, 45.0^\circ]$ & $[6.5^\circ, 9.6^\circ]$ \\
  $2\sigma$ range & $[7.16, 7.99]$ & $[2.17, 2.57]$ & $[31.6^\circ, 35.8^\circ]$
  & $[36.9^\circ, 50.8^\circ]$ & $[5.1^\circ, 10.9^\circ]$ \\
  $3\sigma$ range & $[6.99, 8.18]$ & $[2.06, 2.67]$ & $[30.6^\circ, 36.8^\circ]$
  & $[35.7^\circ, 53.1^\circ]$ & $[1.8^\circ, 12.1^\circ]$ \\
  \hline
\end{tabular}
\end{center}
\end{table}

\section{Two-zero Textures of $M^{}_\nu$}

\subsection{Stability of texture zeros}

We first examine the stability of texture zeros of $M^{}_\nu$
against the one-loop quantum corrections. To be explicit, we
consider the unique dimension-5 Weinberg operator of massive
Majorana neutrinos in an effective field theory after the heavy
degrees of freedom are integrated out \cite{Weinberg}:
\begin{equation}
\frac{{\cal L}^{}_{\rm d =5}}{\Lambda} = \frac{1}{2}
\kappa^{}_{\alpha \beta} \overline{\ell^{}_{\alpha \rm L}} \tilde{H}
\tilde{H}^T \ell^c_{\beta \rm L} + {\rm h.c.} \; ,
\end{equation}
where $\Lambda$ is the cutoff scale, $\ell^{}_{\rm L}$ denotes the
left-handed lepton doublet, $\tilde{H} \equiv i\sigma^{}_2 H^*$ with
$H$ being the standard-model Higgs doublet, and $\kappa$ stands for
the effective neutrino coupling matrix. After spontaneous gauge
symmetry breaking, $\tilde{H}$ gains its vacuum expectation value
$\langle \tilde{H} \rangle = v/\sqrt{2}$ with $v \approx 246$ GeV.
We are then left with the effective Majorana mass matrix $M^{}_\nu =
\kappa v^2/2$ for three light neutrinos from Eq. (19). If the
dimension-5 Weinberg operator is obtained in the framework of the
minimal supersymmetric standard model, one will be left with
$M^{}_\nu = \kappa (v \sin\beta)^2/2$, where $\tan\beta$ denotes the
ratio of the vacuum expectation values of two Higgs doublets. Eq.
(19) or its supersymmetric counterpart can provide a way of
generating tiny neutrino masses. There are a number of interesting
possibilities of building renormalizable gauge models to realize the
effective Weinberg mass operator, such as the well-known seesaw
mechanisms at a superhigh energy scale $\Lambda$ \cite{SS1,SS2,SS3}.

The running of $M^{}_\nu$ from $\Lambda$ to the electroweak scale
$\mu \simeq M^{}_Z$ (or vice versa) is described by the
renormalization-group equations (RGEs) \cite{RGE}. In the chosen
flavor basis and at the one-loop level, $M^{}_\nu (M^{}_Z)$ and
$M^{}_\nu (\Lambda)$ are related to each other via
\begin{eqnarray}
M^{}_\nu (M^{}_Z) = I^{}_0 \left(\matrix{ I^{}_e & 0 & 0 \cr 0 &
I^{}_\mu & 0 \cr 0 & 0 & I^{}_\tau \cr} \right) M^{}_\nu (\Lambda)
\left(\matrix{ I^{}_e & 0 & 0 \cr 0 & I^{}_\mu & 0 \cr 0 & 0 &
I^{}_\tau \cr} \right) \; ,
\end{eqnarray}
where the RGE evolution function $I^{}_0$ denotes the overall
contribution from gauge and quark Yukawa couplings, and
$I^{}_\alpha$ (for $\alpha = e, \mu, \tau$) stand for the
contributions from charged-lepton Yukawa couplings \cite{Mei}.
Because of $I^{}_e < I^{}_\mu < I^{}_\tau$ as a consequence of
$m^{}_e \ll m^{}_\mu \ll m^{}_\tau$, they can modify the texture of
$M^{}_\nu$. In comparison, $I^{}_0 \neq 1$ only affects the overall
mass scale of $M^{}_\nu$. Note, however, that the texture zeros of
$M^{}_\nu$ are stable against such quantum corrections induced by
the one-loop RGEs. Taking Pattern $\bf A^{}_1$ of $M^{}_\nu$ for
example, we have
\begin{eqnarray}
M^{\bf A^{}_1}_\nu (\Lambda) = \left(\matrix{ 0 & 0 & a \cr 0 & b &
c \cr a & c & d \cr} \right)
\end{eqnarray}
at $\Lambda$, and thus
\begin{eqnarray}
M^{\bf A^{}_1}_\nu (M^{}_Z) = I^{}_0 \left(\matrix{ 0 & 0 & a I^{}_e
I^{}_\tau \cr 0 & b I^2_\mu & c I^{}_\mu I^{}_\tau \cr a I^{}_e
I^{}_\tau & c I^{}_\mu I^{}_\tau & d I^2_\tau \cr} \right)
\end{eqnarray}
at $M^{}_Z$. This interesting feature implies that the important
relations obtained in Eqs. (12) and (13) formally hold both at
$\Lambda$ and $M^{}_Z$. In other words, if a seesaw or flavor
symmetry model predicts a two-zero texture of $M^{}_\nu$ at
$\Lambda$, one may simply study its phenomenological consequences at
$M^{}_Z$ by taking account of the same texture zeros. While the
values of neutrino masses and flavor mixing parameters at $M^{}_Z$
turn out to be different from those at $\Lambda$, their correlations
dictated by the texture zeros keep unchanged at any scale between
$\Lambda$ and $M^{}_Z$.

\subsection{Permutation symmetry}

As pointed out in Refs. \cite{FGM}---\cite{More}, the seven viable
two-zero textures of $M^{}_\nu$ can be classified into three
distinct categories: (1) $\bf A^{}_1$ and $\bf A^{}_2$; (2) $\bf
B^{}_1$, $\bf B^{}_2$, $\bf B^{}_3$ and $\bf B^{}_4$; and (3) $\bf
C$. The phenomenological implications of those patterns in the same
category have been found to be almost indistinguishable. Now we show
that there exists a permutation symmetry between Patterns ${\bf
A}^{}_1$ and ${\bf A}^{}_2$, ${\bf B}^{}_1$ and ${\bf B}^{}_2$, or
${\bf B}^{}_3$ and ${\bf B}^{}_4$. This observation may help
understand their similarities in model building and phenomenology.

Let us take Patterns ${\bf A}^{}_1$ and ${\bf A}^{}_2$ in Eq. (2)
for example. Note that the location of texture zeros in ${\bf
A}^{}_1$ can be changed to that in ${\bf A}^{}_2$ by a permutation
in the $2$-$3$ rows and $2$-$3$ columns. To be explicit, we define
the elementary transformation matrix
\begin{eqnarray}
P^{}_{23} = \left(\matrix{1 & 0 & 0 \cr 0 & 0 & 1 \cr 0 & 1 &
0}\right) \; .
\end{eqnarray}
Then the Majorana neutrino mass matrix $M^{\bf A^{}_2}_\nu$ can be
constructed from $M^{\bf A^{}_1}_\nu$ via
\begin{eqnarray}
M^{\bf A^{}_2}_\nu = P^{}_{23} M^{\bf A^{}_1}_\nu P^{\rm T}_{23} \;
.
\end{eqnarray}
If $M^{\bf A^{}_1}_\nu$ is diagonalized by a unitary matrix $V = UP$
like Eq. (8), then Eq. (24) tells us that $M^{\bf A^{}_2}_\nu$ can
be diagonalized by the unitary matrix $\tilde{V} = \tilde{U} P$ with
$\tilde{U} = P^{}_{23} U$. Parametrizing $\tilde{U}$ in terms of
three rotation angles $(\tilde{\theta}^{}_{12},
\tilde{\theta}^{}_{23}, \tilde{\theta}^{}_{13})$ and one
CP-violating phase $\tilde{\delta}$, just like the parametrization
of $U$ in Eq. (9), we immediately obtain  the relations
\begin{eqnarray}
\tilde{\theta}^{}_{12} = \theta^{}_{12} \; , ~~~~
\tilde{\theta}^{}_{13} = \theta^{}_{13} \; , ~~~~
\tilde{\theta}^{}_{23} = \frac{\pi}{2} - \theta^{}_{23} \; , ~~~~
\tilde{\delta} = \delta - \pi \; .
\end{eqnarray}
In addition, $M^{\bf A^{}_1}_\nu$ and $M^{\bf A^{}_2}_\nu$ have the
same eigenvalues $\lambda^{}_i$ (for $i=1,2,3$). The
phenomenological predictions of Pattern $\bf A^{}_2$ can therefore
be obtained from those of Pattern $\bf A^{}_1$ by implementing the
replacements in Eq. (25). It is easy to show that a similar
permutation symmetry holds between Patterns $\bf B^{}_1$ and $\bf
B^{}_2$ or between Patterns $\bf B^{}_3$ and $\bf B^{}_4$.

\subsection{Analytical approximations}

Thanks to the permutation symmetry discussed above, it is only
necessary to study Patterns $\bf A^{}_1$, $\bf B^{}_1$, $\bf B^{}_3$
and $\bf C$ in detail. The analytical results for $\bf A^{}_2$, $\bf
B^{}_2$ and $\bf B^{}_4$ can be obtained, respectively, from those
for $\bf A^{}_1$, $\bf B^{}_1$ and $\bf B^{}_3$ with the
replacements $\theta^{}_{23} \to \pi/2 - \theta^{}_{23}$ and $\delta
\to \delta - \pi$. In this subsection we explore the analytical
relations among three neutrino masses $(m^{}_1, m^{}_2, m^{}_3)$,
three flavor mixing angles $(\theta^{}_{12}, \theta^{}_{23},
\theta^{}_{13})$ and three CP-violating phases $(\delta, \rho,
\sigma)$ for each pattern of $M^{}_\nu$ by means of Eqs.
(11)---(16). The effective mass term of the neutrinoless double-beta
decay, defined as $\langle m \rangle^{}_{ee} \equiv
|(M^{}_\nu)^{}_{ee}|$, will also be discussed.
\begin{itemize}
\item {\bf Pattern $\bf A^{}_1$} with $(M^{}_\nu)^{}_{ee} = (M^{}_\nu)^{}_{e\mu} = 0$.
With the help of Eq. (11), we obtain \cite{xing1}
\begin{eqnarray}
\frac{\lambda^{}_1}{\lambda^{}_3} &=& +\frac{s^{}_{13}}{c^2_{13}}
\left(\frac{s^{}_{12} s^{}_{23}}{c^{}_{12} c^{}_{23}} e^{i\delta}
- s^{}_{13} \right)\; , \nonumber \\
\frac{\lambda^{}_2}{\lambda^{}_3} &=& -\frac{s^{}_{13}}{c^2_{13}}
\left(\frac{c^{}_{12} s^{}_{23}}{s^{}_{12} c^{}_{23}} e^{i\delta} +
s^{}_{13} \right) \; .
\end{eqnarray}
Since $s^2_{13} \ll 1$ is a good approximation, Eq. (26) immediately
leads us to the results
\begin{eqnarray}
\xi = \frac{m^{}_1}{m^{}_3} &\approx& \tan \theta^{}_{12} \tan
\theta^{}_{23} \sin \theta^{}_{13} \; , \nonumber \\
\zeta = \frac{m^{}_2}{m^{}_3} &\approx& \cot \theta^{}_{12} \tan
\theta^{}_{23} \sin \theta^{}_{13} \; ;
\end{eqnarray}
and
\begin{eqnarray}
\rho &\approx& \frac{\delta}{2} \; , \nonumber \\
\sigma &\approx & \frac{\delta}{2} - \frac{\pi}{2} \; .
\end{eqnarray}
Without loss of generality, we choose three neutrino mixing angles
to lie in the first quadrat and allow three CP-violating phases to
vary in the ranges $\rho \in [-\pi/2, +\pi/2]$, $\sigma \in [-\pi/2,
+\pi/2]$ and $\delta \in [0,2\pi]$. Given $0.60 \leq \tan
\theta^{}_{12} \leq 0.75$, $0.72 \leq \tan \theta^{}_{23} \leq 1.3$
and $0.03 \leq \sin \theta^{}_{13} \leq 0.21$ at the $3\sigma$
level, Eq. (27) yields $\xi < \zeta < 1$. So the normal neutrino
mass hierarchy $m^{}_1 < m^{}_2 < m^{}_3$ follows and $\Delta m^2 >
0$ holds. As indicated by Eq. (15), the Dirac CP-violating phase
$\delta$ can completely be fixed if three neutrino mixing angles and
two neutrino mass-squared differences are known. In the lowest order
of $s^{}_{13}$, however,
\begin{equation}
R^{}_\nu = \frac{\delta m^2}{\Delta m^2} \approx \frac{4\tan^2
\theta^{}_{23} \sin^2 \theta^{}_{13}}{\sin 2\theta^{}_{12} \tan
2\theta^{}_{12}} \; ,
\end{equation}
which is actually independent of $\delta$. So we have to go beyond
the leading order approximation. To the next-to-leading order
we obtain
\begin{eqnarray}
\xi = \frac{m^{}_1}{m^{}_3} &\approx& \tan \theta^{}_{12} \tan
\theta^{}_{23} \sin \theta^{}_{13} \sqrt{1 - 2\cot \theta^{}_{12}
\cot \theta^{}_{23} \sin \theta^{}_{13} \cos \delta} \; , \nonumber \\
\zeta = \frac{m^{}_2}{m^{}_3} &\approx& \cot \theta^{}_{12} \tan
\theta^{}_{23} \sin \theta^{}_{13} \sqrt{1 + 2\tan \theta^{}_{12}
\cot \theta^{}_{23} \sin \theta^{}_{13} \cos \delta}\; ;
\end{eqnarray}
and
\begin{equation}
R^{}_\nu \approx \zeta^2 - \xi^2 = \frac{4\tan^2 \theta^{}_{23}
\sin^2 \theta^{}_{13}}{\sin 2\theta^{}_{12} \tan 2\theta^{}_{12}}
\left(1 + \tan 2\theta^{}_{12} \cot \theta^{}_{23} \sin
\theta^{}_{13} \cos \delta\right) \; .
\end{equation}
Finally we arrive at
\begin{equation}
\delta \approx \cos^{-1} \left[ + \frac{\tan \theta^{}_{23}}{\tan
2\theta^{}_{12} \sin \theta^{}_{13}} \left(\frac{\sin
2\theta^{}_{12} \tan 2\theta^{}_{12} R^{}_\nu}{4\tan^2
\theta^{}_{23} \sin^2 \theta^{}_{13}}  - 1\right)\right] \; .
\end{equation}
Taking the best-fit values of the three neutrino mixing angles
(i.e., $\theta^{}_{12} = 33.6^\circ$, $\theta^{}_{23} = 40.4^\circ$
and $\theta^{}_{13} = 8.3^\circ$) together with those of two
neutrino mass-squared differences (i.e., $\delta m^2 = 7.58\times
10^{-5}~{\rm eV}^2$ and $\Delta m^2 = 2.35\times 10^{-3}~{\rm
eV}^2$) from Table 1, one immediately obtains the neutrino mass
spectrum
\footnote{Note again that such best-fit values have been obtained in
the assumption of $\cos \delta = \pm 1$ \cite{Fogli}. To give a
rough estimate, however, we assume that they are essentially
unchanged even for an arbitrary value of $\delta$.}
\begin{eqnarray}
m^{}_3 &\approx& \sqrt{\Delta m^2} = 4.8\times 10^{-2}~{\rm eV} \; ,
\nonumber \\
m^{}_2 &\approx& m^{}_3 \cot \theta^{}_{12} \tan \theta^{}_{23} \sin
\theta^{}_{13} = 8.9\times 10^{-3}~{\rm eV} \; , \nonumber \\
m^{}_1 &\approx& m^{}_3 \tan \theta^{}_{12} \tan \theta^{}_{23} \sin
\theta^{}_{13} = 3.9\times 10^{-3}~{\rm eV} \; ;
\end{eqnarray}
and the CP-violating phases $\delta \approx 64^\circ$, $\rho \approx
32^\circ$ and $\sigma \approx - 58^\circ$. Since $(M^{}_\nu)^{}_{ee}
= 0$ holds for Pattern $\bf A^{}_1$ of $M^{}_\nu$, the effective
mass $\langle m \rangle^{}_{ee}$ of the neutrinoless double-beta
decay is definitely vanishing.

\item {\bf Pattern $\bf A^{}_2$} with $(M^{}_\nu)^{}_{ee} =
(M^{}_\nu)^{}_{e\tau} = 0$. As pointed out in section 3.2, all the
analytical results of Pattern ${\bf A}^{}_2$ can be obtained from
those of Pattern $\bf A^{}_1$ with the replacements $\theta^{}_{23}
\to \pi/2 - \theta^{}_{23}$ and $\delta \to \delta -\pi$. So it is
straightforward to have
\begin{eqnarray}
\xi = \frac{m^{}_1}{m^{}_3} &\approx& \tan \theta^{}_{12} \cot
\theta^{}_{23} \sin \theta^{}_{13} \sqrt{1 + 2\cot \theta^{}_{12}
\tan \theta^{}_{23} \sin \theta^{}_{13} \cos \delta} \; , \nonumber \\
\zeta = \frac{m^{}_2}{m^{}_3} &\approx& \cot \theta^{}_{12} \cot
\theta^{}_{23} \sin \theta^{}_{13} \sqrt{1 - 2\tan \theta^{}_{12}
\tan \theta^{}_{23} \sin \theta^{}_{13} \cos \delta}\; ;
\end{eqnarray}
and
\begin{equation}
\delta \approx \cos^{-1} \left[-\frac{\cot \theta^{}_{23}}{\tan
2\theta^{}_{12} \sin \theta^{}_{13}} \left(\frac{\sin
2\theta^{}_{12} \tan 2\theta^{}_{12} R^{}_\nu}{4\cot^2
\theta^{}_{23} \sin^2 \theta^{}_{13}}  - 1\right)\right] \; .
\end{equation}
In addition,
\begin{eqnarray}
\rho &\approx& \frac{\delta}{2} - \frac{\pi}{2} \; , \nonumber \\
\sigma &\approx & \frac{\delta}{2}  \; ;
\end{eqnarray}
and the neutrino mass spectrum turns out to be
\begin{eqnarray} m^{}_3 &\approx& \sqrt{\Delta m^2} \; ,
\nonumber \\
m^{}_2 &\approx& m^{}_3 \cot \theta^{}_{12} \cot \theta^{}_{23} \sin
\theta^{}_{13} \; , \nonumber \\
m^{}_1 &\approx& m^{}_3 \tan \theta^{}_{12} \cot \theta^{}_{23} \sin
\theta^{}_{13} \; .
\end{eqnarray}
Similar to the case of Pattern $\bf A^{}_1$, $\langle m
\rangle^{}_{ee} = |(M^{}_\nu)^{}_{ee}| = 0$ holds in Pattern $\bf
A^{}_2$. Provided the same best-fit values of $\delta m^2$, $\Delta
m^2$, $\theta^{}_{12}$, $\theta^{}_{23}$ and $\theta^{}_{13}$ are
taken, one can easily verify that $\cos \delta
> 1$ holds, implying a potential inconsistency of Pattern $\bf A^{}_2$ with
current experimental data. When the uncertainties of those neutrino
mixing parameters are taken into account (e.g., at the $3\sigma$
level), however, we actually find no problem with Pattern $\bf
A^{}_2$ (see the numerical analysis in section 4).

\item {\bf Pattern $\bf B^{}_1$} with $(M^{}_\nu)^{}_{\mu\mu} =
(M^{}_\nu)^{}_{e\tau} = 0$. With the help of Eq. (11), we obtain
\begin{eqnarray}
\frac{\lambda^{}_1}{\lambda^{}_3} &=& \frac{s^{}_{12} c^{}_{12}
s^{}_{23} (2c^2_{23} s^2_{13} - s^2_{23} c^2_{13}) - c^{}_{23}
s^{}_{13} (s^2_{12} s^2_{23} e^{+i\delta} + c^2_{12} c^2_{23}
e^{-i\delta})}{s^{}_{12} c^{}_{12} s^{}_{23} c^2_{23} + (s^2_{12} -
c^2_{12})c^3_{23} s^{}_{13} e^{i\delta} + s^{}_{12} c^{}_{12}
s^{}_{23} s^2_{13} (1+c^2_{23})e^{2i\delta}} e^{2i\delta} \; ,
\nonumber \\
\frac{\lambda^{}_2}{\lambda^{}_3} &=& \frac{s^{}_{12} c^{}_{12}
s^{}_{23} (2c^2_{23} s^2_{13} - s^2_{23} c^2_{13}) + c^{}_{23}
s^{}_{13} (c^2_{12} s^2_{23} e^{+i\delta} + s^2_{12} c^2_{23}
e^{-i\delta})}{s^{}_{12} c^{}_{12} s^{}_{23} c^2_{23} + (s^2_{12} -
c^2_{12})c^3_{23} s^{}_{13} e^{i\delta} + s^{}_{12} c^{}_{12}
s^{}_{23} s^2_{13} (1+c^2_{23})e^{2i\delta}} e^{2i\delta} \; .
\end{eqnarray}
In the leading order approximation,
\begin{eqnarray}
\xi = \frac{m^{}_1}{m^{}_3} &\approx& \tan^2 \theta^{}_{23} \; ,
\nonumber \\
\zeta = \frac{m^{}_2}{m^{}_3} &\approx& \tan^2 \theta^{}_{23} \; ;
\end{eqnarray}
and
\begin{eqnarray}
\rho \approx \sigma \approx \delta - \frac{\pi}{2} \; .
\end{eqnarray}
In the next-to-leading order approximation, we find
\begin{eqnarray}
&& \frac{m^{}_1}{m^{}_3} - \frac{m^{}_2}{m^{}_3} \approx +\frac{4\sin
\theta^{}_{13} \cos \delta}{\sin 2\theta^{}_{12} \sin
2\theta^{}_{23}} \; ,
\nonumber \\
&& \rho - \sigma \approx - \frac{2\sin
\theta^{}_{13} \sin \delta}{\sin 2\theta^{}_{12} \tan
2\theta^{}_{23} \tan^2 \theta^{}_{23}} \; .
\end{eqnarray}
The Dirac CP-violating phase $\delta$ can be determined from
\begin{equation}
R^{}_\nu \approx \frac{2\sin \theta^{}_{13}}{\sin 2\theta^{}_{12}}
\left|\tan 2\theta^{}_{23} \cos \delta\right| \; .
\end{equation}
Since $\delta m^2 > 0$ or equivalently $m^{}_2 > m^{}_1$, we have
$\cos \delta < 0$ from Eq. (41). Taking the best-fit values, one can
obtain $\delta \approx 91^\circ$ and thus $\rho \approx \sigma
\approx 1^\circ$. Eq. (41) tells us that the difference between
$m^{}_2/m^{}_3$ and $m^{}_1/m^{}_3$ is about 0.01, and that between
$\rho$ and $\sigma$ is about $4^\circ$. On the other hand, the
neutrino mass spectrum is given by
\begin{eqnarray}
&& m^{}_3 \approx \sqrt{\frac{\Delta m^2}{1 - \tan^4
\theta^{}_{23}}} \approx 7.0\times 10^{-2}~{\rm eV} \; ,
\nonumber \\
&& m^{}_2 \approx m^{}_1
\approx m^{}_3 \tan^2 \theta^{}_{23} \approx 5.1\times 10^{-2}~{\rm
eV} \; ;
\end{eqnarray}
and the effective mass term of the neutrinoless double-beta decay turns
out to be $\langle m \rangle^{}_{ee} \approx
m^{}_3 \tan^2 \theta^{}_{23} \approx 5.1\times 10^{-2}~{\rm eV}$.
Note that we have input the best-fit value $\theta^{}_{23} =
40.4^\circ$ \cite{Fogli},
so the normal neutrino mass hierarchy $m^{}_3 > m^{}_2 > m^{}_1$
is allowed. If the $3\sigma$ range of $\theta^{}_{23}$ is
taken, however, both normal and inverted mass hierarchies are likely.
When $\tan \theta^{}_{23} > 1$, which is allowed at the $2\sigma$
or $3\sigma$ level \cite{Fogli}, we have the inverted neutrino
mass hierarchy with $\Delta m^2 < 0$. In short,
the analytical formulas for $m^{}_1$,
$m^{}_2$ and $m^{}_3$ in Eq. (43) are valid no matter which mass
hierarchy is taken, but the corresponding numerical results depend
on the input value of $\theta^{}_{23}$.

\item {\bf Pattern $\bf B^{}_2$} with $(M^{}_\nu)^{}_{\tau\tau} =
(M^{}_\nu)^{}_{e\mu} = 0$. By using the permutation symmetry
discussed in section 3.2, we obtain
\begin{eqnarray}
\xi = \frac{m^{}_1}{m^{}_3} &\approx& \cot^2 \theta^{}_{23} \; ,
\nonumber \\
\zeta = \frac{m^{}_2}{m^{}_3} &\approx& \cot^2 \theta^{}_{23} \; ;
\end{eqnarray}
and
\begin{eqnarray}
\rho \approx \sigma \approx \delta - \frac{\pi}{2} \;
\end{eqnarray}
in the leading order approximation. Note that there is a period of
$\pi$ for $\rho$ and $\sigma$, so Eq. (45) is identical to Eq. (40)
even if the replacement $\delta \to \delta - \pi$ is made. In the
next-to-leading order approximation, we have
\begin{eqnarray}
&& \frac{m^{}_1}{m^{}_3} - \frac{m^{}_2}{m^{}_3} \approx
-\frac{4\sin \theta^{}_{13} \cos \delta}{\sin 2\theta^{}_{12} \sin
2\theta^{}_{23}} \; , \nonumber \\
&& \rho - \sigma \approx - \frac{2\sin \theta^{}_{13} \sin
\delta}{\sin 2\theta^{}_{12} \tan 2\theta^{}_{23} \cot^2
\theta^{}_{23}} \; .
\end{eqnarray}
The Dirac CP-violating phase is also determined by Eq. (42), because
$|\tan 2\theta^{}_{23}|$ keeps invariant under the replacement
$\theta^{}_{23} \to \pi/2 - \theta^{}_{23}$. Now we have $\cos
\delta > 0$ in view of the fact of $m^{}_2 > m^{}_1$. So $\delta
\approx 89^\circ$ and $\rho \approx \sigma \approx -1^\circ$. Eq.
(46) tells us that the difference between $m^{}_2/m^{}_3$ and
$m^{}_1/m^{}_3$ is about 0.01, and that between $\rho$ and $\sigma$
is about $2^\circ$. The neutrino mass spectrum turns out to be
\begin{eqnarray}
&& m^{}_3 \approx \sqrt{\frac{\Delta m^2}{1 - \cot^4
\theta^{}_{23}}} \approx 5.1\times 10^{-2}~{\rm eV} \; ,
\nonumber \\
&& m^{}_2 \approx m^{}_1 \approx m^{}_3 \cot^2 \theta^{}_{23}
\approx 7.0\times 10^{-2}~{\rm eV} \; .
\end{eqnarray}
In addition, $\langle m \rangle^{}_{ee} \approx m^{}_3 \cot^2
\theta^{}_{23} \approx 7.0\times 10^{-2}~{\rm eV}$. Note that we
have input the best-fit value $\theta^{}_{23} = 40.4^\circ$
\cite{Fogli}, so the inverted mass hierarchy $m^{}_2 > m^{}_1
> m^{}_3$ appears. If the $3\sigma$ range of $\theta^{}_{23}$ is
taken, however, both normal and inverted mass hierarchies are
likely. Hence it is experimentally important to determine the
deviation of $\theta^{}_{23}$ from $\pi/4$.

\item {\bf Pattern $\bf B^{}_3$} with $(M^{}_\nu)^{}_{\mu\mu}
= (M^{}_\nu)^{}_{e\mu} = 0$. With the help of Eq. (11), we obtain
\begin{eqnarray}
\frac{\lambda^{}_1}{\lambda^{}_3} &=& -\frac{s^{}_{23}}{c^{}_{23}}
\cdot \frac{s^{}_{12} s^{}_{23} - c^{}_{12} c^{}_{23} s^{}_{13}
e^{-i\delta}}{s^{}_{12} c^{}_{23} + c^{}_{12} s^{}_{23} s^{}_{13}
e^{+i\delta}} e^{2i\delta} \;, \nonumber \\
\frac{\lambda^{}_2}{\lambda^{}_3} &=& -\frac{s^{}_{23}}{c^{}_{23}}
\cdot \frac{c^{}_{12} s^{}_{23} + s^{}_{12} c^{}_{23} s^{}_{13}
e^{-i\delta}}{c^{}_{12} c^{}_{23} - s^{}_{12} s^{}_{23} s^{}_{13}
e^{+i\delta}} e^{2i\delta} \;.
\end{eqnarray}
In the leading order approximation,
\begin{eqnarray}
\xi = \frac{m^{}_1}{m^{}_3} &\approx& \tan^2 \theta^{}_{23} \; ,
\nonumber \\
\zeta = \frac{m^{}_2}{m^{}_3} &\approx& \tan^2 \theta^{}_{23} \; ;
\end{eqnarray}
and
\begin{eqnarray}
\rho \approx \sigma \approx \delta - \frac{\pi}{2} \; .
\end{eqnarray}
In the next-to-leading order approximation, we have
\begin{eqnarray}
&& \frac{m^{}_1}{m^{}_3} - \frac{m^{}_2}{m^{}_3} \approx -\frac{4
\tan^2 \theta^{}_{23} \sin \theta^{}_{13} \cos \delta}{\sin
2\theta^{}_{12} \sin 2\theta^{}_{23}} \; , \nonumber \\
&& \rho - \sigma \approx + \frac{2\sin \theta^{}_{13} \sin
\delta}{\sin 2\theta^{}_{12} \tan 2\theta^{}_{23}} \; .
\end{eqnarray}
The Dirac CP-violating phase $\delta$ can be determined from
\begin{equation}
R^{}_\nu \approx \frac{2\sin \theta^{}_{13}}{\sin 2\theta^{}_{12}}
\tan^2 \theta^{}_{23}\left|\tan 2\theta^{}_{23} \cos \delta\right|
\; .
\end{equation}
The condition $m^{}_2 > m^{}_1$ leads to $\cos \delta > 0$. Taking
the best-fit values \cite{Fogli}, we obtain $\delta \approx
89^\circ$ and $\rho \approx \sigma \approx -1^\circ$ from Eq. (51).
The difference between $m^{}_2/m^{}_3$ and $m^{}_1/m^{}_3$ is about
0.008, and that between $\rho$ and $\sigma$ is about $3^\circ$. On
the other hand, the neutrino mass spectrum is given by
\begin{eqnarray}
&& m^{}_3 \approx \sqrt{\frac{\Delta m^2}{1 - \tan^4
\theta^{}_{23}}} \approx 7.0\times 10^{-2}~{\rm eV} \; ,
\nonumber \\
&& m^{}_2 \approx m^{}_1 \approx m^{}_3 \tan^2 \theta^{}_{23}
\approx 5.1\times 10^{-2}~{\rm eV} \; .
\end{eqnarray}
The effective mass term of the neutrinoless double-beta decay turns
out to be $\langle m \rangle^{}_{ee} \approx m^{}_3 \tan^2
\theta^{}_{23} \approx 5.1\times 10^{-2}~{\rm eV}$. Note that the
phenomenology of Pattern $\bf B^{}_3$ is essentially the same as
that of Pattern $\bf B^{}_1$ except for the Dirac CP-violating phase
$\delta$. Note also that we have used the best-fit value
$\theta^{}_{23} = 40.4^\circ$ \cite{Fogli}, so the normal neutrino
mass hierarchy $m^{}_3
> m^{}_2 > m^{}_1$ is allowed. If the $3\sigma$ range of
$\theta^{}_{23}$ is taken, however, both normal and inverted
neutrino mass hierarchies are likely.

\item {\bf Pattern $\bf B^{}_4$} with $(M^{}_\nu)^{}_{\tau\tau} =
(M^{}_\nu)^{}_{e\tau} = 0$. By using the permutation symmetry
discussed in section 3.2, we obtain
\begin{eqnarray}
\xi = \frac{m^{}_1}{m^{}_3} &\approx& \cot^2 \theta^{}_{23} \; ,
\nonumber \\
\zeta = \frac{m^{}_2}{m^{}_3} &\approx& \cot^2 \theta^{}_{23} \; ;
\end{eqnarray}
and
\begin{eqnarray}
\rho \approx \sigma \approx \delta - \frac{\pi}{2} \;
\end{eqnarray}
in the leading order approximation. More accurately, we have
\begin{eqnarray}
&& \frac{m^{}_1}{m^{}_3} - \frac{m^{}_2}{m^{}_3} \approx
+\frac{4\cot^2 \theta^{}_{23} \sin \theta^{}_{13} \cos \delta}{\sin
2\theta^{}_{12} \sin 2\theta^{}_{23}} \; , \nonumber \\
&& \rho - \sigma \approx + \frac{2\sin \theta^{}_{13} \sin
\delta}{\sin 2\theta^{}_{12} \tan 2\theta^{}_{23}} \; .
\end{eqnarray}
The Dirac CP-violating phase is given by
\begin{eqnarray}
R^{}_\nu \approx \frac{2\sin \theta^{}_{13}}{\sin 2\theta^{}_{12}}
\cot^2 \theta^{}_{23}\left|\tan 2\theta^{}_{23} \cos \delta\right|
\; .
\end{eqnarray}
We find $\cos \delta < 0$ from the fact of $m^{}_2 > m^{}_1$. Hence
$\delta \approx 91^\circ$ and $\rho \approx \sigma \approx 1^\circ$.
Eq. (56) tells us that the difference between $m^{}_2/m^{}_3$ and
$m^{}_1/m^{}_3$ is about 0.015, and that between $\rho$ and $\sigma$
is about $3^\circ$. The neutrino mass spectrum turns out to be
\begin{eqnarray}
&& m^{}_3 \approx \sqrt{\frac{\Delta m^2}{1 - \cot^4
\theta^{}_{23}}} \approx 5.1\times 10^{-2}~{\rm eV} \; , \nonumber
\\
&& m^{}_2 \approx m^{}_1 \approx m^{}_3 \cot^2 \theta^{}_{23}
\approx 7.0\times 10^{-2}~{\rm eV} \; .
\end{eqnarray}
In addition, $\langle m \rangle^{}_{ee} \approx m^{}_3 \cot^2
\theta^{}_{23} \approx 7.0\times 10^{-2}~{\rm eV}$. Note that we
have used the best-fit value $\theta^{}_{23} = 40.4^\circ$
\cite{Fogli}, so the inverted neutrino mass hierarchy $m^{}_2 >
m^{}_1 > m^{}_3$ appears. If the $3\sigma$ range of $\theta^{}_{23}$
is taken, however, both normal and inverted mass hierarchies are
likely. It is obvious that a precise determination of $\delta$ and
the neutrino mass hierarchy is crucial to pin down one of the four
patterns $\bf B^{}_1$, $\bf B^{}_2$, $\bf B^{}_3$ and $\bf B^{}_4$.
All of them predict a nearly degenerate neutrino mass spectrum.

\item {\bf Pattern $\bf C$} with $(M^{}_\nu)^{}_{\mu\mu} =
(M^{}_\nu)^{}_{\tau\tau} = 0$. With the help of Eq. (11), we obtain
\begin{eqnarray}
\frac{\lambda^{}_1}{\lambda^{}_3} &=& \frac{c^{}_{12}
c^2_{13}}{s^{}_{13}} \cdot \frac{-c^{}_{12} (s^2_{23} - c^2_{23}) -
2s^{}_{12} s^{}_{23} c^{}_{23} s^{}_{13} e^{i\delta}}{2s^{}_{12}
c^{}_{12} s^{}_{23} c^{}_{23} - (s^2_{12} - c^2_{12})(s^2_{23} -
c^2_{23})s^{}_{13} e^{i\delta} + 2s^{}_{12} c^{}_{12} s^{}_{23}
c^{}_{23}s^2_{13} e^{2i\delta}} e^{i\delta} \; , ~~~~~~\nonumber \\
\frac{\lambda^{}_2}{\lambda^{}_3} &=& \frac{s^{}_{12}
c^2_{13}}{s^{}_{13}} \cdot \frac{+s^{}_{12} (s^2_{23} - c^2_{23}) -
2c^{}_{12} s^{}_{23} c^{}_{23} s^{}_{13} e^{i\delta}}{2s^{}_{12}
c^{}_{12} s^{}_{23} c^{}_{23} - (s^2_{12} - c^2_{12})(s^2_{23} -
c^2_{23})s^{}_{13} e^{i\delta} + 2s^{}_{12} c^{}_{12} s^{}_{23}
c^{}_{23} s^2_{13} e^{2i\delta}} e^{i\delta} \; .
\end{eqnarray}
To the lowest order,
\begin{eqnarray}
\xi = \frac{m^{}_1}{m^{}_3} &\approx& \sqrt{1 - \frac{2 \cot
\theta^{}_{12} \cos \delta}{\tan 2\theta^{}_{23} \sin
\theta^{}_{13}} + \frac{\cot^2 \theta^{}_{12}}{\tan^2
2\theta^{}_{23} \sin^2 \theta^{}_{13}}} \; , \nonumber \\
\zeta = \frac{m^{}_2}{m^{}_3} &\approx& \sqrt{1 + \frac{2 \tan
\theta^{}_{12} \cos \delta}{\tan 2\theta^{}_{23} \sin
\theta^{}_{13}} + \frac{\tan^2 \theta^{}_{12}}{\tan^2
2\theta^{}_{23} \sin^2 \theta^{}_{13}}} \; .
\end{eqnarray}
Since $m^{}_2 > m^{}_1$, Eq. (60) implies $\tan 2\theta^{}_{23} \cos
\delta > 0$ and $m^{}_2 > m^{}_3$. Furthermore, one can verify that
$\tan 2\theta^{}_{12} \tan 2\theta^{}_{23} \sin \theta^{}_{13} \cos
\delta > 1$ is required to guarantee $m^{}_2 > m^{}_1$. Hence only
the inverted mass hierarchy $m^{}_2
> m^{}_1 > m^{}_3$ is allowed. The Dirac CP-violating phase $\delta$
can be determined from
\begin{equation}
R^{}_\nu \approx \frac{2\left(1+\tan \theta^{}_{12} \tan
\theta^{}_{23}\right) \left(\tan 2\theta^{}_{12} \tan
2\theta^{}_{23} \sin \theta^{}_{13} \cos \delta - 1\right)}{\tan
\theta^{}_{12} \tan \theta^{}_{23} \tan 2\theta^{}_{12} \left(\tan
\theta^{}_{12} + 2\tan 2\theta^{}_{23} \sin \theta^{}_{13} \cos
\delta\right)} \; .
\end{equation}
Taking the best-fit values of three neutrino mixing angles and two
neutrino mass-squared differences, we obtain $\delta \approx
61^\circ$ from Eq. (61). The Majorana CP-violating phases $\rho$ and
$\sigma$ turn out to be
\begin{eqnarray}
\rho &\approx& \delta + \frac{1}{2} \tan^{-1}\left[\frac{\cot
\theta^{}_{12} \sin \delta}{\tan 2\theta^{}_{23} \sin \theta^{}_{13}
- \cot \theta^{}_{12} \cos \delta}\right] - \frac{\pi}{2} \approx
+13^\circ
\;, \nonumber \\
\sigma &\approx& \delta - \frac{1}{2} \tan^{-1}\left[\frac{\tan
\theta^{}_{12} \sin \delta}{\tan 2\theta^{}_{23} \sin \theta^{}_{13}
+ \tan \theta^{}_{12} \cos \delta}\right] - \frac{\pi}{2} \approx
-42^\circ \; .
\end{eqnarray}
Finally we obtain
\begin{eqnarray}
m^{}_3 &\approx& \sqrt{\frac{\tan^2 2\theta^{}_{23}
\cot^2\theta^{}_{12} \sin^2 \theta^{}_{13} \Delta m^2}{1 + 2\cot
\theta^{}_{12} \tan 2\theta^{}_{23} \sin \theta^{}_{13} \cos
\delta}} \approx 4.3\times 10^{-2}~{\rm eV} \; , \nonumber \\
m^{}_2 &\approx& m^{}_3 \sqrt{1 + \frac{2 \tan \theta^{}_{12} \cos
\delta}{\tan 2\theta^{}_{23} \sin \theta^{}_{13}} + \frac{\tan^2
\theta^{}_{12}}{\tan^2 2\theta^{}_{23} \sin^2 \theta^{}_{13}}}
\approx 6.6\times 10^{-2}~{\rm eV} \; , \nonumber \\
m^{}_1 &\approx& m^{}_3 \sqrt{1 - \frac{2 \cot \theta^{}_{12} \cos
\delta}{\tan 2\theta^{}_{23} \sin \theta^{}_{13}} + \frac{\cot^2
\theta^{}_{12}}{\tan^2 2\theta^{}_{23} \sin^2 \theta^{}_{13}}}
\approx 6.5\times 10^{-2}~{\rm eV} \;
\end{eqnarray}
together with
\begin{equation}
\langle m \rangle^{}_{ee} \approx m^{}_3 \sqrt{1- \frac{4 \cot
2\theta^{}_{12} \cos \delta}{\tan 2\theta^{}_{23} \sin
\theta^{}_{13}} + \frac{4\cot^2 2\theta^{}_{12}}{\tan^2
2\theta^{}_{23} \sin^2 \theta^{}_{13}}} \approx 4.1\times
10^{-2}~{\rm eV} \; .
\end{equation}
Note that both $\tan 2\theta^{}_{23}
> 0$ with $\cos \delta > 0$ and $\tan 2\theta^{}_{23} < 0$ with
$\cos \delta < 0$ are likely, if the $3\sigma$ range of
$\theta^{}_{23}$ is taken into account. Nevertheless, the inverted
mass hierarchy $m^{}_3 < m^{}_1 < m^{}_2$ is expected in both cases.
Going beyond the above analytical approximation, we find
that the normal mass hierarchy $m^{}_1 < m^{}_2 < m^{}_3$ is actually 
allowed if and only if $\theta^{}_{23}$ lies in the vicinity
of $45^\circ$
\footnote{We thank the anonymous referee for pointing out this
interesting possibility to us.}.
This point can be understood in a simple way. With the help of Eq. (59), 
one obtains 
\begin{eqnarray}
\frac{\lambda^{}_1}{\lambda^{}_3} =
\frac{\lambda^{}_2}{\lambda^{}_3} = -\frac{c^2_{13}e^{2i\delta}}
{1+s^2_{13} e^{2i\delta}} \approx - c^2_{13} e^{2i\delta}
\left( 1 - s^2_{13} e^{2i\delta} \right) \;
\end{eqnarray}
in the limit of $\theta^{}_{23} = 45^\circ$, implying the mass
hierarchy $m^{}_1 = m^{}_2 < m^{}_3$ for $\theta^{}_{13} \neq 0^\circ$. 
Hence a tiny deviation of
$\theta^{}_{23}$ from $45^\circ$ is required to lift the degeneracy
of $m^{}_1$ and $m^{}_2$ and then reproduce the small
but nonvanishing value of $R^{}_\nu$, such that 
$m^{}_1 < m^{}_2 < m^{}_3$ comes out. Because the parameter space
of this possibility is too small, we shall
mainly concentrate on the inverted mass hierarchy in our subsequent
numerical analysis.
\end{itemize}
In summary, the analytical results for the three CP-violating phases
$(\delta, \rho, \sigma)$, the two neutrino mass ratios
$m^{}_1/m^{}_3$ and $m^{}_2/m^{}_3$, the absolute neutrino mass
$m^{}_3$ and the effective neutrino mass $\langle m
\rangle^{}_{ee}$, predicted by seven two-zero textures of
$M^{}_\nu$, are listed in Tables 2---4. If the best-fit values of
neutrino mixing parameters \cite{Fogli} are taken, we find that only
Pattern $\bf A^{}_2$ can be excluded. We expect that all the seven
patterns of $M^{}_\nu$ can survive current experimental tests at the
$3\sigma$ level. A detailed numerical analysis will be done in
section 4.

\section{Numerical Analysis}

The analytical results obtained above show that the neutrino mass
hierarchy is actually related to the flavor mixing angle
$\theta^{}_{23}$ in Patterns $\bf B^{}_1$, $\bf B^{}_2$, $\bf
B^{}_3$ and $\bf B^{}_4$ of $M^{}_\nu$. In particular, it depends on
whether $\theta^{}_{23} > 45^\circ$ or $\theta^{}_{23} < 45^\circ$.
According to Table 1 \cite{Fogli}, only $\theta^{}_{23} \leq
45^\circ$ is allowed at the $1\sigma$ level
\footnote{Note that $\theta^{}_{23} \geq 45^\circ$ seems to be
favored at the $1\sigma$ level in the global analysis done by
Schwetz {\it et al} \cite{Schwetz}.}.
If a two-zero texture of $M^{}_\nu$ can accommodate both normal and
inverted mass hierarchies, we shall only concentrate on the one
dictated by $\theta^{}_{23} < 45^\circ$ in our numerical analysis,
because the other possibility is just an opposite and trivial
exercise. We have noticed that the Dirac CP-violating phase $\delta$
should be close to $\pi/2$ or $3\pi/2$ in Patterns $\bf
B^{}_{1,2,3,4}$, and the differences between two Majorana
CP-violating phases $\rho$ and $\sigma$ in these two cases are
distinct. For illustration, we shall only focus on the range of
$\delta$ around $\pi/2$ in our numerical analysis of these four
patterns. The explicit calculations are done in the following way:
\begin{enumerate}
\item For each of the seven patterns of $M^{}_\nu$
we generate a set of random numbers of $(\theta^{}_{12},
\theta^{}_{23}, \theta^{}_{13})$ and $(\delta m^2, \Delta m^2)$
lying in their $3\sigma$ ranges given by Eqs. (17) and (18) together
with a random value of $\delta$ in the range $\delta \in [0,2\pi]$.

\item Given the above random numbers, it is possible to
calculate other physical parameters of $M^{}_\nu$, including three
neutrino mass eigenvalues $(m^{}_1, m^{}_2, m^{}_3)$, two
Majorana-type CP-violating phases $(\rho, \sigma)$, the effective
mass of the neutrinoless double-beta decay $\langle m
\rangle^{}_{ee}$ and the Jarlskog invariant of leptonic CP violation
$J^{}_{\rm CP} = s^{}_{12} c^{}_{12} s^{}_{23} c^{}_{23} s^{}_{13}
c^2_{13} \sin\delta$. To judge whether a pattern of $M^{}_\nu$ is
consistent with current experimental data or not, we require that
the consistency conditions should be satisfied: (a) because of
$\delta m^2 > 0$, we require $m^{}_2 > m^{}_1$ or equivalently
$\zeta^2 - \xi^2 > 0$; (b) since only the neutrino mass hierarchies
$m^{}_2 > m^{}_1 > m^{}_3$ and $m^{}_3 > m^{}_2 > m^{}_1$ are
phenomenologically allowed, we further require $(\zeta^2 - 1)(\xi^2
- 1) > 0$ (i.e., $\xi^2 > 1$ and $\zeta^2
> 1$ correspond to the inverted mass hierarchy, whereas
$\xi^2 < 1$ and $\zeta^2 < 1$ stand for the normal mass hierarchy);
(c) given the values of three neutrino mixing angles and two
mass-squared differences, the Dirac CP-violating phase $\delta$ is
actually fixed by Eq. (15). Instead of solving $\delta$, we require
that Eq. (15) should be satisfied up to a reasonable degree of
precision (e.g., $10^{-4}$).

\item We consider all the points satisfying the consistency
conditions, and then have a nine-dimensional parameter space spanned
by nine quantities $(\theta^{}_{12}, \theta^{}_{23}, \theta^{}_{13},
\delta, \rho, \sigma, m^{}_1, m^{}_2, m^{}_3)$. The low-energy
observables such as $J^{}_{\rm CP}$ and $\langle m \rangle^{}_{ee}$
can accordingly be calculated. To present the final results in a
simple and clear way, we restrict ourselves to the two-dimensional
parameter space and set the $x$-axis to be the Dirac CP-violating
phase $\delta$. Therefore, what we actually show are the allowed
ranges of relevant physical parameters of each pattern of
$M^{}_\nu$.

\item Corresponding to the allowed ranges of $\theta^{}_{12}$,
$\theta^{}_{23}$ and $\theta^{}_{13}$ changing with $\delta$, their
histograms are also plotted because they can signify the most
probable values of three flavor mixing angles. The height of each
histogram indicates the number of points in each bin. Since the
future neutrino oscillation experiments will greatly improve the
measurements of three neutrino mixing angles, such a presentation
should be helpful in judging which two-zero pattern of $M^{}_\nu$ is
phenomenologically more favored.
\end{enumerate}
We stress that the strategy of our numerical analysis can also apply
to other textures of the Majorana neutrino mass matrix $M^{}_\nu$.
The two-zero patterns under discussion will serve as a good example
to illustrate this strategy.

Our numerical results are presented in Figs. 1---14. Comments and
discussions follow.
\begin{itemize}
\item {\bf Pattern $\bf A^{}_1$.} Fig. 1 tells us that the neutrino
mixing angles, in particular $\theta^{}_{12}$ and $\theta^{}_{23}$,
are actually insensitive to the Dirac CP-violating phase $\delta$.
This point can easily be understood with the help of Eq. (31) or Eq.
(32), in which $\delta$ is only loosely related to the ratio of two
neutrino mass-squared differences $R^{}_\nu$ due to the smallness of
$\theta^{}_{13}$. An interesting observation from the left panel of
Fig. 1 is, that a relatively large value of $\theta^{}_{13}$ (i.e.,
$\theta^{}_{13} \approx 7^\circ \cdots 8^\circ$), which is quite
close to the best-fit value $\theta^{}_{13} = 8.3^\circ$
\cite{Fogli}, is particularly favored. On the other hand, the
neutrino mass spectrum is weakly hierarchical, as shown in Fig. 2.
The dependence of $m^{}_1$ and $m^{}_2$ on $\delta$ is ascribed to
the next-to-leading order corrections given in Eq. (30). We have
also illustrated the numerical prediction for $J^{}_{\rm CP}$ in
Fig. 2. One can see the maximal value of $J^{}_{\rm CP}$ is at the
percent level and should be able to lead to observable effects of CP
violation in a variety of long-baseline neutrino oscillation
experiments. Because $\delta$ itself is essentially unconstrained by
current experimental data at the $3\sigma$ level, $\rho$ and
$\sigma$ turn out to be arbitrary as shown in Fig. 2, although their
correlations with $\delta$ are rather sharp. Finally we remark that
Pattern $\bf A^{}_1$ of $M^{}_\nu$ predicts $\langle
m\rangle^{}_{ee} = 0$ for the neutrinoless double-beta decay.

\item {\bf Pattern $\bf A^{}_2$.} As shown in Figs. 3 and 4, the
phenomenological implications of Pattern $\bf A^{}_2$ of $M^{}_\nu$
are essentially the same as those of Pattern $\bf A^{}_1$. For
instance, $\theta^{}_{13} \approx 6^\circ \cdots 8^\circ$ is favored
and the effective mass term $\langle m\rangle^{}_{ee}$ is vanishing.
Thus it is only necessary to emphasize their main difference. We
have demonstrated a permutation symmetry between $M^{\bf
A^{}_1}_\nu$ and $M^{\bf A^{}_2}_\nu$ in section 3.2 and found that
the present best-fit values of neutrino mixing parameters (mainly
$\theta^{}_{23} = 40.4^\circ$ \cite{Fogli}) cannot coincide with
Pattern $\bf A^{}_2$. If the maximal mixing angle $\theta^{}_{23} =
45^\circ$ were finally established, however, it would be almost
impossible to distinguish between Patterns $\bf A^{}_1$ and $\bf
A^{}_2$ in practice.

\item {\bf Pattern $\bf B^{}_1$.} Since $m^{}_2 > m^{}_1$, we have
$\cos \delta < 0$ (i.e., $\pi/2 < \delta < 3\pi/2$) in this case. In
our numerical analysis we only focus on the range $\delta \in
[\pi/2, \pi]$, because the range $\delta \in [\pi, 3\pi/2]$ can
similarly be analyzed. As shown in Figs. 5 and 6, only a very narrow
region $\delta \in [0.50\pi, 0.56\pi]$ is phenomenologically
allowed. This result obviously originates from Eq. (42), where
$|\cos \delta|$ must be small enough to suppress the magnitude of
$R^{}_\nu$. Furthermore, we only consider the normal neutrino mass
hierarchy corresponding to $\theta^{}_{23} < 45^\circ$. The latter
seems to be favored by current data at the $1\sigma$ level
\cite{Fogli}. Fig. 5 shows that $\theta^{}_{13} \sim 3^\circ$ and
$\theta^{}_{23} \sim 37^\circ$ are more likely. A strong correlation
between $\theta^{}_{23}$ and $\delta$ can also be understood from
Eq. (42); namely, the maximal mixing $\theta^{}_{23} \approx
45^\circ$ requires the maximal CP-violating phase $\delta \approx
\pi/2$. In addition, a nearly degenerate mass spectrum as shown in
Fig. 6 is predicted. There is a lower bound on $\langle m
\rangle^{}_{ee}$ (i.e., $\langle m \rangle^{}_{ee} \geq 0.03~{\rm
eV}$), which will be tested in the future experiments of the
neutrinoless double-beta decay. Note that $\langle m \rangle^{}_{ee}
\approx 0.2~{\rm eV}$ can be achieved when $\theta^{}_{23} \approx
45^\circ$ and $\delta \approx \pi/2$. The other three patterns of
this category (i.e., $\bf B^{}_2$, $\bf B^{}_3$ and $\bf B^{}_4$)
have similar consequences, for which the numerical results have been
given in Figs. 7 and 8, Figs. 9 and 10, and Figs. 11 and 12,
respectively. But the details of these patterns, such as the allowed
ranges of $(\rho, \sigma, \delta)$ and $(m^{}_1, m^{}_2, m^{}_3)$,
are somewhat different.

\item {\bf Pattern ${\bf C}$.} Fig. 13 shows no significant
preference in the allowed ranges of three neutrino mixing angles. A
strong correlation between $\theta^{}_{23}$ and $\delta$ only
appears when $\delta$ is close to $\pi/2$. As shown in Fig. 14 and
discussed in section 3.3, the inverted neutrino mass hierarchy
is allowed in most parts of the parameter space (and a 
normal mass hierarchy is possible only when $\theta^{}_{23}$
is very close to $45^\circ$ and $\theta^{}_{13}$ is nonvanishing). 
Like Patterns $\bf B^{}_{1,2,3,4}$, there is a lower bound on $\langle m
\rangle^{}_{ee}$ in Pattern $\bf C$ (i.e., $\langle m
\rangle^{}_{ee} > 0.02~{\rm eV}$), and its maximal value can
saturate the present experimental upper bound $\langle m
\rangle^{}_{ee} < 0.3~{\rm eV}$ \cite{2Beta}. It should be noted
that the sum of the three neutrino masses is subject to the
cosmological bound $\sum m^{}_i < 0.58~{\rm eV}$ at the $95\%$
confidence level \cite{WMAP}, which has been derived from the
seven-year WMAP data on the cosmic background radiation combined
with the Baryon Acoustic Oscillations. Therefore, the possibility of
$\delta \sim \pi/2$ is essentially excluded at the same confidence
level. Similar conclusions apply to Patterns $\bf B^{}_{1,2,3,4}$.
\end{itemize}
Finally it is worth pointing out that we have also done a numerical
analysis of the two-zero textures of $M^{}_\nu$ by using the
$1\sigma$ and $2\sigma$ values of $\delta m^2$, $\Delta m^2$,
$\theta^{}_{12}$, $\theta^{}_{23}$ and $\theta^{}_{13}$ (see Table 1
and Ref. \cite{Fogli}). We find that all the seven patterns
discussed above are compatible with current experimental data at the
$1\sigma$ or $2\sigma$ level, although the corresponding parameter
space is somewhat smaller. To be conservative, we take our numerical
results obtained at the $3\sigma$ level more seriously.

\section{Texture Zeros from Flavor Symmetries}

In general, the texture zeros of a Majorana neutrino mass matrix can
be realized in various seesaw models with proper discrete flavor
symmetries. It is possible to obtain the zeros in arbitrary entries
of a fermion mass matrix by means of the Abelian symmetries (e.g.,
the cyclic group $Z^{}_n$ \cite{Grimus1}). To illustrate how to
realize the two-zero textures of $M^{}_\nu$ discussed above, we
shall work in the type-II seesaw model, which extends the scalar
sector of the standard model with one or more $SU(2)^{}_{\rm L}$
scalar triplets \cite{SS2}. For $N$ scalar triplets, the
gauge-invariant Lagrangian relevant for neutrino masses reads
\begin{equation}
-{\cal L}^{}_{\Delta} = \frac{1}{2} \sum_{j} \sum_{\alpha, \beta}
\left(Y^{}_{\Delta^{}_j}\right)^{}_{\alpha \beta}
\overline{\ell^{}_{\alpha \rm L}} \Delta^{}_j i\sigma^{}_2
\ell^c_{\beta \rm L} + {\rm h.c.} \; ,
\end{equation}
where $\alpha$ and $\beta$ run over $e$, $\mu$ and $\tau$,
$\Delta^{}_j$ denotes the $j$-th triplet scalar field (for $j = 1,
2, \cdots, N$), and $Y^{}_{\Delta^{}_j}$ is the corresponding Yukawa
coupling matrix. After the triplet scalar acquires its vacuum
expectation value $\langle \Delta^{}_j\rangle \equiv
v^{}_{\Delta^{}_j}$, the Majorana neutrino mass matrix is given by
\begin{equation}
M^{}_\nu = \sum_j Y^{}_{\Delta^{}_j} v^{}_{\Delta^{}_j} \; ,
\end{equation}
where the smallness of $v^{}_{\Delta^{}_j}$ is attributed to the
largeness of the triplet scalar mass scale \cite{SS2}.

In order to generate the texture zeros in $M^{}_\nu$ and derive the
seven viable patterns ${\bf A^{}_{1,2}}$, ${\bf B^{}_{1,2,3,4}}$ and
${\bf C}$, we follow the spirit of Ref. \cite{Grimus2} and impose
the $Z^{}_n$ symmetry on the Lagrangian in Eq. (66). The unique
generator of the cyclic group $Z^{}_n$ is $\varpi = e^{i2\pi/n}$,
which produces all the group elements $Z^{}_n = \{{\bf 1}, \varpi,
\varpi^2, \cdots, \varpi^{n-1}\}$. Now it is straightforward to
specify the number of scalar triplets $N$ (i.e., $j = 1, 2, \cdots,
N$), the order of the cyclic group $n$ and the representations of
$\ell^{}_\alpha$ and $\Delta^{}_j$ under the symmetry group.
\begin{itemize}
\item $N = 3$ and $n = 6$ for ${\bf A^{}_{1,2}}$ and ${\bf
B^{}_{3,4}}$. In this case we have to introduce three scalar
triplets and the $Z^{}_6$ symmetry group, for which the generator is
$\varpi = e^{i\pi/3}$. The representations of lepton doublets
$\ell^{}_{\alpha {\rm L}}$ (for $\alpha = e, \mu, \tau$) and the
scalar triplets $\Delta^{}_j$ (for $j = 1, 2, 3$) are assigned as
follows.
\begin{enumerate}

\item For Pattern ${\bf A^{}_{1}}$:
\begin{eqnarray}
&& \ell^{}_{e{\rm L}} \to \varpi^2 \ell^{}_{e{\rm L}} \; , ~~~
\ell^{}_{\mu{\rm L}} \to {\bf 1} ~\ell^{}_{\mu{\rm L}} \; , ~~~
\ell^{}_{\tau{\rm L}} \to \varpi^3 \ell^{}_{\tau{\rm L}} \; ,
\nonumber \\
&& \Delta^{}_1 \to {\bf 1} ~\Delta^{}_1 \; , ~~~~ \Delta^{}_2 \to
\varpi^3 \Delta^{}_2 \; , ~~~ \Delta^{}_3 \to \varpi \Delta^{}_3 \;
.
\end{eqnarray}
Given the above representations of lepton doublets, three scalar
triplets are needed to enforce the nonzero elements in the neutrino
mass matrix: $\Delta^{}_1$ for $(M^{}_\nu)^{}_{\mu\mu}$ and
$(M^{}_\nu)^{}_{\tau\tau}$, $\Delta^{}_2$ for
$(M^{}_\nu)^{}_{\mu\tau}$ and $\Delta^{}_3$ for
$(M^{}_\nu)^{}_{e\tau}$.

\item For Pattern ${\bf A^{}_{2}}$:
\begin{eqnarray}
&& \ell^{}_{e{\rm L}} \to \varpi^2 \ell^{}_{e{\rm L}} \; , ~~~
\ell^{}_{\mu{\rm L}} \to {\bf 1} ~\ell^{}_{\mu{\rm L}} \; , ~~~
\ell^{}_{\tau{\rm L}} \to \varpi^3 \ell^{}_{\tau{\rm L}} \; ,
\nonumber \\
&& \Delta^{}_1 \to {\bf 1} ~\Delta^{}_1 \; , ~~~~ \Delta^{}_2 \to
\varpi^3 \Delta^{}_2 \; , ~~~ \Delta^{}_3 \to \varpi^4 \Delta^{}_3
\; .
\end{eqnarray}
Note that Eq. (69) differs from Eq. (68) only in the assignment for
the triplet $\Delta^{}_3$. Such a difference originates from the
fact that $(M^{}_\nu)^{}_{e\mu} = 0$ and $(M^{}_\nu)^{}_{e\tau} \neq
0$ hold for Pattern ${\bf A^{}_1}$, while $(M^{}_\nu)^{}_{e\tau} =
0$ and $(M^{}_\nu)^{}_{e\mu} \neq 0$ hold for Pattern ${\bf
A^{}_2}$. It is worthwhile to point out that the assignments in Eq.
(69) are by no means unique. As we have discussed in section 3.2,
there exists a permutation symmetry between ${\bf A^{}_{1}}$ and
${\bf A^{}_{2}}$. Therefore, we may exchange the representations of
$\ell^{}_{\mu{\rm L}}$ and $\ell^{}_{\tau{\rm L}}$ in Eq. (68) but
preserve those of scalar triplets to obtain $\bf A^{}_2$ from $\bf
A^{}_1$.

\item For Pattern ${\bf B^{}_{3}}$:
\begin{eqnarray}
&& \ell^{}_{e{\rm L}} \to {\bf 1} ~\ell^{}_{e{\rm L}} \; , ~~~
\ell^{}_{\mu{\rm L}} \to \varpi^2 \ell^{}_{\mu{\rm L}} \; , ~~~
\ell^{}_{\tau{\rm L}} \to \varpi^3 \ell^{}_{\tau{\rm L}} \; ,
\nonumber \\
&& \Delta^{}_1 \to {\bf 1} ~\Delta^{}_1 \; , ~~~ \Delta^{}_2 \to
\varpi^3 \Delta^{}_2 \; , ~~~~ \Delta^{}_3 \to \varpi \Delta^{}_3 \;
.
\end{eqnarray}

\item For Pattern ${\bf B^{}_{4}}$:
\begin{eqnarray}
&& \ell^{}_{e{\rm L}} \to \varpi^3 \ell^{}_{e{\rm L}} \; , ~~~
\ell^{}_{\mu{\rm L}} \to {\bf 1} ~\ell^{}_{\mu{\rm L}} \; , ~~~
\ell^{}_{\tau{\rm L}} \to \varpi^2 \ell^{}_{\tau{\rm L}} \; ,
\nonumber \\
&& \Delta^{}_1 \to {\bf 1} ~\Delta^{}_1 \; , ~~~~ \Delta^{}_2 \to
\varpi^3 \Delta^{}_2 \; , ~~~ \Delta^{}_3 \to \varpi^4 \Delta^{}_3
\; .
\end{eqnarray}
In addition to the permutation symmetry $P^{}_{23}$ discussed in
section 3.2, we notice that the location of texture zeros in Pattern
${\bf B^{}_3}$ can be obtained from that in Pattern ${\bf A^{}_1}$
by a permutation in the $1$-$2$ rows and $1$-$2$ columns. Similarly
there exists a $1$-$3$ permutation symmetry between Pattern ${\bf
B^{}_4}$ and Pattern ${\bf A^{}_2}$. Although these symmetries
cannot lead to simple relations between any two of the neutrino
mixing parameters, they are instructive for the assignments of
lepton doublets. For instance, Eq. (70) and Eq. (71) can be derived
from Eq. (68) and Eq. (69) by exchanging the representations of
$\ell^{}_{e{\rm L}}$ and $\ell^{}_{\mu{\rm L}}$ and the
representations of $\ell^{}_{e{\rm L}}$ and $\ell^{}_{\tau{\rm L}}$,
respectively. The assignments of scalar triplets should not be
changed.
\end{enumerate}

\item $N = 2$ and $n = 3$ for ${\bf B^{}_{1,2}}$. In this case we
introduce only two scalar triplets and the $Z^{}_3$ symmetry group,
for which the generator is $\varpi = e^{2i\pi/3}$. The assignments
of lepton doublets $\ell^{}_{\alpha {\rm L}}$ (for $\alpha = e, \mu,
\tau$) and the scalar triplets $\Delta^{}_j$ (for $j = 1, 2$) are as
follows.

\begin{enumerate}
\item For Pattern ${\bf B^{}_{1}}$:
\begin{eqnarray}
\ell^{}_{e{\rm L}} \to {\bf 1} ~\ell^{}_{e{\rm L}} \; , ~~
\ell^{}_{\mu{\rm L}} \to \varpi \ell^{}_{\mu{\rm L}} \; , ~~
\ell^{}_{\tau{\rm L}} \to \varpi^2 \ell^{}_{\tau{\rm L}} \; , ~~
\Delta^{}_1 \to {\bf 1} ~\Delta^{}_1 \; , ~~ \Delta^{}_2 \to
\varpi^2 \Delta^{}_2 \; .
\end{eqnarray}

\item For Pattern ${\bf B^{}_{2}}$:
\begin{eqnarray}
\ell^{}_{e{\rm L}} \to {\bf 1} ~\ell^{}_{e{\rm L}} \; , ~~
\ell^{}_{\mu{\rm L}} \to \varpi^2 \ell^{}_{\mu{\rm L}} \; , ~~
\ell^{}_{\tau{\rm L}} \to \varpi \ell^{}_{\tau{\rm L}} \; , ~~
\Delta^{}_1 \to {\bf 1} ~\Delta^{}_1 \; , ~~ \Delta^{}_2 \to
\varpi^2 \Delta^{}_2 \; .
\end{eqnarray}
Note that the assignment in Eq. (73) is slightly different from that
in Ref. \cite{Grimus2}, where $\Delta^{}_2 \to \varpi \Delta^{}_2$
is taken and the representations of lepton doublets are the same as
in Eq. (72). Here we have implemented the $2$-$3$ permutation
symmetry between Pattern ${\bf B^{}_{1}}$ and Pattern ${\bf
B^{}_{2}}$.
\end{enumerate}

\item $N = 3$ and $n = 4$ for Pattern ${\bf C}$.  In this case we have to
introduce three scalar triplets and the $Z^{}_4$ symmetry group, for
which the generator is $\varpi = e^{i\pi/2}$. The assignments of
lepton doublets $\ell^{}_{\alpha {\rm L}}$ (for $\alpha = e, \mu,
\tau$) and the scalar triplets $\Delta^{}_j$ (for $j = 1, 2, 3$) are
\begin{eqnarray}
&& \ell^{}_{e{\rm L}} \to {\bf 1} ~\ell^{}_{e{\rm L}} \; , ~~~
\ell^{}_{\mu{\rm L}} \to \varpi \ell^{}_{\mu{\rm L}} \; , ~~~
\ell^{}_{\tau{\rm L}} \to \varpi^3 \ell^{}_{\tau{\rm L}} \; ,
\nonumber \\
&& \Delta^{}_1 \to {\bf 1} ~\Delta^{}_1 \; , ~~~ \Delta^{}_2 \to
\varpi^3 \Delta^{}_2 \; , ~~~ \Delta^{}_3 \to \varpi \Delta^{}_3 \;
.
\end{eqnarray}
\end{itemize}
Thus all the seven two-zero patterns of $M^{}_\nu$ can be obtained
in this simple symmetry scheme.

In all cases we have taken the right-handed charged-lepton
singlets $E^{}_{\alpha {\rm R}}$ to transform in the same way as the
left-handed lepton doublets $\ell^{}_{\alpha {\rm L}}$ and taken the
standard-model Higgs doublet $H$ to be in the trivial
representation. Hence the charged-lepton mass matrix $M^{}_l$
is diagonal, as we have chosen from the beginning. The two-zero
textures of $M^{}_\nu$ can also be realized in the seesaw models
with three right-handed neutrinos, several Higgs singlets, doublets
and triplets, by imposing either Abelian or non-Abelian discrete
flavor symmetries \cite{Grimus1,Grimus2,symmetry}.

\section{Summary}

In view of the latest T2K and MINOS neutrino oscillation data which
hint at a relatively large value of $\theta^{}_{13}$, we have
performed a systematic study of the Majorana neutrino mass matrix
$M^{}_\nu$ with two independent texture zeros. It turns out that
seven patterns (i.e., ${\bf A^{}_{1,2}}$, ${\bf B^{}_{1,2,3,4}}$ and
${\bf C}$) can survive current experimental tests at the $3\sigma$
level, although they are also compatible with the data at the
$1\sigma$ or $2\sigma$ level. The following is a brief summary of
our main observations:
\begin{itemize}
\item Given the values of three flavor mixing angles
$(\theta^{}_{12}, \theta^{}_{23}, \theta^{}_{13})$ and two neutrino
mass-squared differences $(\delta m^2, \Delta m^2)$, it is in
principle possible to fully determine three CP-violating phases
$(\delta, \rho, \sigma)$ and three neutrino masses $(m^{}_1, m^{}_2,
m^{}_3)$. The analytical formulas for the latter have been derived
and listed in Tables 2---4.

\item By making the analytical approximations and taking the best-fit
values of neutrino mixing parameters \cite{Fogli}, we find that only
Pattern ${\bf A^{}_2}$ can be excluded. We have numerically
confirmed that all the seven patterns of $M^{}_\nu$ (i.e., ${\bf
A^{}_{1,2}}$, ${\bf B^{}_{1,2,3,4}}$ and ${\bf C}$) are compatible
with current neutrino oscillation data at the $1\sigma$ level, but
our numerical results have been presented only at the more
conservative $3\sigma$ level.

\item Figs. 1---14 show the main numerical results of our systematic
analysis. Some interesting points should be emphasized. (1) Both
${\bf A^{}_1}$ and ${\bf A^{}_2}$ favor a relatively large
$\theta^{}_{13}$ (e.g., $\theta^{}_{13} \sim 8^\circ$), ${\bf
B^{}_{1,2,3,4}}$ prefer a relatively small $\theta^{}_{13}$ (e.g.,
$\theta^{}_{13} \sim 3^\circ$), and ${\bf C}$ shows no significant
preference for the magnitude of $\theta^{}_{13}$. (2) The Dirac
CP-violating phase $\delta$ obtained from ${\bf B^{}_{1,2,3,4}}$
lies in a narrow range around $\pi/2$ or $3\pi/2$, and $\delta =
\pi/2$ is strongly correlated with $\theta^{}_{23} = \pi/4$. (3) For
$\delta \to \pi/2$ and $\theta^{}_{23} \to \pi/4$, the predictions
for neutrino masses $m^{}_i$ and the effective neutrino mass
$\langle m \rangle^{}_{ee}$ may run into contradiction with their
upper bounds set by the cosmological observations and the
neutrinoless double-beta decay experiments. (4) The size of
$J^{}_{\rm CP}$ may reach the percent level and thus appreciable
leptonic CP violation is possible to show up in the future
long-baseline neutrino oscillation experiments.
\end{itemize}
In addition we have shown that the texture zeros of the Majorana
neutrino mass matrix $M^{}_\nu$ are stable against the one-loop
quantum corrections
\footnote{In contrast, the texture zeros of the Dirac neutrino mass
matrix are essentially sensitive to quantum corrections like those
of quark mass matrices \cite{Rode}.},
and pointed out that there exists a permutation symmetry between
${\bf A}^{}_1$ and ${\bf A}^{}_2$, ${\bf B}^{}_1$ and ${\bf B}^{}_2$
or ${\bf B}^{}_3$ and ${\bf B}^{}_4$. In the type-II seesaw model
with two or three scalar triplets we have illustrated how to realize
two-zero textures of $M^{}_\nu$ by using the $Z^{}_n$ flavor
symmetry.

The ongoing and upcoming neutrino oscillation experiments are
expected to measure the neutrino mixing parameters, in particular
the smallest mixing angle $\theta^{}_{13}$, the deviation of
$\theta^{}_{23}$ from $\pi/4$ and the Dirac CP-violating phase
$\delta$. The sensitivity of future cosmological observations to the
sum of neutrino masses $\sum m^{}_i$ and the sensitivity of the
neutrinoless double-beta decay experiments to the effective mass
term $\langle m \rangle^{}_{ee}$ will probably reach $\sim0.05~{\rm
eV}$ in the near future. We therefore expect that some patterns of
the two-zero textures of $M^{}_\nu$ might be excluded or only
marginally allowed by tomorrow's data, and those capable of
surviving should shed light on the underlying flavor structures of
massive neutrinos.

\vspace{0.5cm}
\begin{flushleft}
{\large \bf Acknowledgements}
\end{flushleft}

This work was supported in part by the National Natural Science
Foundation of China under grant No. 10875131 (Z.Z.X.) and by the
Alexander von Humboldt Foundation (S.Z.).

\newpage

\begin{table}[t]
\renewcommand\arraystretch{2.1}
\caption{Seven viable patterns of the neutrino mass matrix
$M^{}_\nu$ with two texture zeros, and their predictions for three
CP-violating phases $(\delta, \rho, \sigma)$.}
\begin{center}
\begin{tabular}{cccc}
\hline
  \hline
  Pattern & Texture of $M^{}_\nu$ &~& CP-violating phases \\
  \hline
  $\bf A^{}_1$ & $\left(\matrix{{0} & {0} & \times \cr
                                {0} & \times & \times \cr
                                \times & \times & \times \cr}\right)$
                                &~&
                                $\begin{array}{c}
                                \displaystyle \delta \approx \cos^{-1} \left[+\frac{\tan \theta^{}_{23}}{\tan
                                2\theta^{}_{12} \sin \theta^{}_{13}} \left(\frac{\sin
                                2\theta^{}_{12} \tan 2\theta^{}_{12} R^{}_\nu}{4\tan^2
                                \theta^{}_{23} \sin^2 \theta^{}_{13}}  - 1\right)\right] \\
                                \displaystyle \rho \approx
                                \frac{\delta}{2} \; ,~~  \sigma \approx \frac{\delta}{2} -
                                \frac{\pi}{2}
                                \end{array}$ \\
                                \hline
  $\bf A^{}_2$ & $\left(\matrix{{0} & \times & {0} \cr
                                \times & \times & \times \cr
                                {0} & \times & \times \cr}\right)$
                                &~&
                                $\begin{array}{c}
                                \displaystyle \delta \approx \cos^{-1} \left[-\frac{\cot \theta^{}_{23}}{\tan
                                2\theta^{}_{12} \sin \theta^{}_{13}} \left(\frac{\sin
                                2\theta^{}_{12} \tan 2\theta^{}_{12} R^{}_\nu}{4\cot^2
                                \theta^{}_{23} \sin^2 \theta^{}_{13}}  - 1\right)\right] \\
                                \displaystyle \rho \approx
                                \frac{\delta}{2} - \frac{\pi}{2} \; ,~~  \sigma \approx \frac{\delta}{2}
                                \end{array}$ \\
                                \hline
  $\bf B^{}_1$ & $\left(\matrix{\times & \times & {0} \cr
                                \times & {0} & \times \cr
                                {0} & \times & \times \cr}\right)$
                                &~&
                                $\begin{array}{c} \displaystyle \delta \approx \cos^{-1}\left[-
                                \frac{\sin 2\theta^{}_{12} R^{}_\nu}{2\sin \theta^{}_{13}
                                |\tan 2\theta^{}_{23}|}\right] \\
                                \displaystyle \rho \approx \sigma \approx \delta - \frac{\pi}{2} \; , ~~
                                \rho - \sigma \approx - \frac{2\sin \theta^{}_{13} \sin \delta}
                                {\sin 2\theta^{}_{12} \tan 2\theta^{}_{23} \tan^2 \theta^{}_{23}}
                                \end{array}$ \\
                                \hline
  $\bf B^{}_2$ & $\left(\matrix{\times & {0} & \times \cr
                                {0} & \times & \times \cr
                                \times & \times & {0} \cr}\right)$ &~&
                                $\begin{array}{c} \displaystyle \delta \approx
                                \cos^{-1}\left[+
                                \frac{\sin 2\theta^{}_{12} R^{}_\nu}{2\sin \theta^{}_{13}
                                |\tan 2\theta^{}_{23}|}\right] \\
                                \displaystyle \rho \approx \sigma \approx \delta - \frac{\pi}{2} \; , ~~
                                \rho - \sigma \approx - \frac{2\sin \theta^{}_{13} \sin \delta}
                                {\sin 2\theta^{}_{12} \tan 2\theta^{}_{23} \cot^2 \theta^{}_{23}} \end{array}$ \\
                                \hline
  $\bf B^{}_3$ & $\left(\matrix{\times & {0} & \times \cr
                                {0} & {0} & \times \cr
                                \times & \times & \times \cr}\right)$ & ~&
                                $\begin{array}{c} \displaystyle \delta \approx
                                \cos^{-1}\left[+
                                \frac{\sin 2\theta^{}_{12} \cot^2 \theta^{}_{23} R^{}_\nu}{2\sin \theta^{}_{13}
                                |\tan 2\theta^{}_{23}|}\right] \\
                                \displaystyle \rho \approx \sigma \approx \delta - \frac{\pi}{2} \; , ~~
                                \rho - \sigma \approx + \frac{2\sin \theta^{}_{13} \sin \delta}
                                {\sin 2\theta^{}_{12} \tan 2\theta^{}_{23}} \end{array}$ \\
                                \hline
  $\bf B^{}_4$ & $\left(\matrix{\times & \times & {0} \cr
                                \times & \times & \times \cr
                                {0} & \times & {0} \cr}\right)$ &~&
                                $\begin{array}{c} \displaystyle \delta \approx \cos^{-1}\left[-
                                \frac{\sin 2\theta^{}_{12} \tan^2 \theta^{}_{23} R^{}_\nu}{2\sin \theta^{}_{13}
                                |\tan 2\theta^{}_{23}|}\right] \\
                                \displaystyle \rho \approx \sigma \approx \delta - \frac{\pi}{2} \; , ~~
                                \rho - \sigma \approx + \frac{2\sin \theta^{}_{13} \sin \delta}
                                {\sin 2\theta^{}_{12} \tan 2\theta^{}_{23}} \end{array}$ \\
                                \hline
  $\bf C$ & $\left(\matrix{\times & \times & \times \cr
                                \times & {0} & \times \cr
                                \times & \times & {0} \cr}\right)$ &~&
                                $\begin{array}{c}
                                \displaystyle \delta \approx \frac{2(1+\tan \theta^{}_{12} \tan \theta^{}_{23})
                                + \tan^2 \theta^{}_{12} \tan 2\theta^{}_{12} \tan \theta^{}_{23}
                                R^{}_\nu}{\left[1+(1-R^{}_\nu)\tan \theta^{}_{12} \tan \theta^{}_{23}
                                \right]\tan 2\theta^{}_{12} \tan 2\theta^{}_{23} \sin \theta^{}_{13}} \\
                                \displaystyle \rho \approx \delta + \frac{1}{2} \tan^{-1}\left[ \frac{\cot \theta^{}_{12}
                                \sin \delta}{\tan 2\theta^{}_{23} \sin \theta^{}_{13} - \cot \theta^{}_{12}
                                \cos \delta}\right] - \frac{\pi}{2} \\
                                \displaystyle \sigma \approx \delta - \frac{1}{2} \tan^{-1}\left[ \frac{\tan \theta^{}_{12}
                                \sin \delta}{\tan 2\theta^{}_{23} \sin \theta^{}_{13} + \tan \theta^{}_{12}
                                \cos \delta}\right] - \frac{\pi}{2}\end{array}$ \\
  \hline\hline
\end{tabular}
\end{center}
\end{table}

\begin{table}[t]
\caption{Seven viable patterns of the neutrino mass matrix
$M^{}_\nu$ with two texture zeros, and their predictions for two
neutrino mass ratios $\xi \equiv m^{}_1/m^{}_3$ and $\zeta \equiv
m^{}_2/m^{}_3$.}
\begin{center}
\begin{tabular}{ccccc}
\hline
  \hline
  Pattern &~& Texture of $M^{}_\nu$ &~& Neutrino mass ratios \\
  \hline
  $\bf A^{}_1$ &~& $\left(\matrix{{0} & {0} & \times \cr
                                {0} & \times & \times \cr
                                \times & \times & \times \cr}\right)$
                                &~&
                                $\begin{array}{l}
                                \xi \approx \tan \theta^{}_{12} \tan \theta^{}_{23} \sin
                                \theta^{}_{13}\; , ~~
                                \zeta \approx \cot \theta^{}_{12} \tan \theta^{}_{23} \sin \theta^{}_{13}\\
                                \end{array}$ \\
                                \hline
  $\bf A^{}_2$ &~& $\left(\matrix{{0} & \times & {0} \cr
                                \times & \times & \times \cr
                                {0} & \times & \times \cr}\right)$
                                &~&
                                $\begin{array}{l}
                                \xi \approx \tan \theta^{}_{12} \cot \theta^{}_{23} \sin \theta^{}_{13}
                                \; , ~~
                                \zeta \approx \cot \theta^{}_{12} \cot \theta^{}_{23} \sin \theta^{}_{13}\\
                                \end{array}$ \\
                                \hline
  $\bf B^{}_1$ &~& $\left(\matrix{\times & \times & {0} \cr
                                \times & {0} & \times \cr
                                {0} & \times & \times \cr}\right)$
                                &~&
                                $\begin{array}{l}\xi \approx \zeta \approx \tan^2
                                \theta^{}_{23}\; ,~~~
                                \displaystyle \xi - \zeta \approx + \frac{4 \sin \theta^{}_{13}
                                \cos\delta}{\sin 2\theta^{}_{12} \sin 2\theta^{}_{23}} \end{array}$ \\
                                \hline
  $\bf B^{}_2$ &~& $\left(\matrix{\times & {0} & \times \cr
                                {0} & \times & \times \cr
                                \times & \times & {0} \cr}\right)$ &~&
                                $\begin{array}{l}\xi \approx \zeta \approx \cot^2
                                \theta^{}_{23}\; ,~~~
                                \displaystyle \xi - \zeta \approx - \frac{4 \sin \theta^{}_{13}
                                \cos\delta}{\sin 2\theta^{}_{12} \sin 2\theta^{}_{23}} \end{array}$ \\
                                \hline
  $\bf B^{}_3$ &~& $\left(\matrix{\times & {0} & \times \cr
                                {0} & {0} & \times \cr
                                \times & \times & \times \cr}\right)$ & ~&
                                $\begin{array}{l}\xi \approx \zeta \approx \tan^2
                                \theta^{}_{23}\; ,~~~
                                \displaystyle \xi - \zeta \approx - \frac{4 \tan^2 \theta^{}_{23}
                                \sin \theta^{}_{13} \cos\delta}{\sin 2\theta^{}_{12} \sin 2\theta^{}_{23}} \end{array}$ \\
                                \hline
  $\bf B^{}_4$ &~& $\left(\matrix{\times & \times & {0} \cr
                                \times & \times & \times \cr
                                {0} & \times & {0} \cr}\right)$ &~&
                                $\begin{array}{l}\xi \approx \zeta \approx \cot^2
                                \theta^{}_{23} \; ,~~~
                                \displaystyle \xi - \zeta \approx + \frac{4 \cot^2 \theta^{}_{23}
                                \sin \theta^{}_{13} \cos\delta}{\sin 2\theta^{}_{12} \sin 2\theta^{}_{23}} \end{array}$ \\
                                \hline
  $\bf C$ &~& $\left(\matrix{\times & \times & \times \cr
                                \times & {0} & \times \cr
                                \times & \times & {0} \cr}\right)$ &~&
                                $\begin{array}{l}
                                \displaystyle \xi \approx \left(1 - \frac{2\cot \theta^{}_{12}\cos \delta}{ \tan 2\theta^{}_{23}
                                \sin \theta^{}_{13}} + \frac{\cot^2 \theta^{}_{12} }{\tan^2 2\theta^{}_{23}
                                \sin^2 \theta^{}_{13}}\right)^{1/2} \\
                                \displaystyle \zeta \approx \left(1 + \frac{2\tan \theta^{}_{12}\cos \delta}{ \tan 2\theta^{}_{23}
                                \sin \theta^{}_{13}} + \frac{\tan^2 \theta^{}_{12} }{\tan^2 2\theta^{}_{23}
                                \sin^2 \theta^{}_{13}}\right)^{1/2} \end{array}$ \\
  \hline\hline
\end{tabular}
\end{center}
\end{table}

\begin{table}[t]
\renewcommand\arraystretch{2.1}
\caption{Seven viable patterns of the neutrino mass matrix
$M^{}_\nu$ with two texture zeros, and their predictions for the
absolute neutrino mass $m^{}_3$ and the effective mass terms of the
neutrinoless double-beta decay $\langle m \rangle^{}_{ee}$.}
\begin{center}
\begin{tabular}{cccc}
  \hline
  \hline
  Pattern & Texture of $M^{}_\nu$ &~& The scales of neutrino masses \\
  \hline
  $\bf A^{}_1$ & $\left(\matrix{{0} & {0} & \times \cr
                                {0} & \times & \times \cr
                                \times & \times & \times \cr}\right)$
                                &~&
                                $\begin{array}{c}
                                m^{}_3 \approx \sqrt{\Delta m^2} \;
                                ,~~~ \langle m \rangle^{}_{ee} = 0
                                \end{array}$ \\
                                \hline
  $\bf A^{}_2$ & $\left(\matrix{{0} & \times & {0} \cr
                                \times & \times & \times \cr
                                {0} & \times & \times \cr}\right)$
                                &~&
                                $\begin{array}{c}
                                m^{}_3 \approx \sqrt{\Delta m^2} \;
                                ,~~~ \langle m \rangle^{}_{ee} = 0
                                \end{array}$ \\
                                \hline
  $\bf B^{}_1$ & $\left(\matrix{\times & \times & {0} \cr
                                \times & {0} & \times \cr
                                {0} & \times & \times \cr}\right)$
                                &~&
                                $\begin{array}{c} \displaystyle m^{}_3 \approx \sqrt{
                                \frac{\Delta m^2}{1 - \tan^4 \theta^{}_{23}}} \; , ~~
                                \langle m \rangle^{}_{ee} \approx m^{}_3 \tan^2 \theta^{}_{23} \end{array}$ \\
                                \hline
  $\bf B^{}_2$ & $\left(\matrix{\times & {0} & \times \cr
                                {0} & \times & \times \cr
                                \times & \times & {0} \cr}\right)$ &~&
                                $\begin{array}{l} \displaystyle m^{}_3 \approx \sqrt{
                                \frac{\Delta m^2}{1 - \cot^4 \theta^{}_{23}}} \; , ~~
                                \langle m \rangle^{}_{ee} \approx m^{}_3 \cot^2 \theta^{}_{23} \end{array}$ \\
                                \hline
  $\bf B^{}_3$ & $\left(\matrix{\times & {0} & \times \cr
                                {0} & {0} & \times \cr
                                \times & \times & \times \cr}\right)$ & ~&
                                $\begin{array}{c} \displaystyle m^{}_3 \approx \sqrt{
                                \frac{\Delta m^2}{1 - \tan^4 \theta^{}_{23}}} \; , ~~
                                \langle m \rangle^{}_{ee} \approx m^{}_3 \tan^2 \theta^{}_{23} \end{array}$ \\
                                \hline
  $\bf B^{}_4$ & $\left(\matrix{\times & \times & {0} \cr
                                \times & \times & \times \cr
                                {0} & \times & {0} \cr}\right)$ &~&
                                $\begin{array}{c} \displaystyle m^{}_3 \approx \sqrt{
                                \frac{\Delta m^2}{1 - \cot^4 \theta^{}_{23}}} \; , ~~
                                \langle m \rangle^{}_{ee} \approx m^{}_3 \cot^2 \theta^{}_{23} \end{array}$ \\
                                \hline
  $\bf C$ & $\left(\matrix{\times & \times & \times \cr
                                \times & {0} & \times \cr
                                \times & \times & {0} \cr}\right)$ &~&
                                $\begin{array}{c}
                                \displaystyle m^{}_3 \approx \sqrt{\frac{\tan^2 2\theta^{}_{23}
                                \cot^2 \theta^{}_{12} \sin^2\theta^{}_{13} \Delta m^2}
                                {1+2\cot\theta^{}_{12} \tan 2\theta^{}_{23} \sin\theta^{}_{13}
                                \cos \delta}} \\
                                \displaystyle \langle m \rangle^{}_{ee} \approx m^{}_3
                                \sqrt{1 - \frac{4\cot 2\theta^{}_{12} \cos\delta}{\tan 2\theta^{}_{23}
                                \sin \theta^{}_{13}} + \frac{4\cot^2 2\theta^{}_{12}}{\tan^2 2\theta^{}_{23}
                                \sin^2 \theta^{}_{13}}}\end{array}$ \\
  \hline\hline
\end{tabular}
\end{center}
\end{table}

\newpage

\begin{figure}[b]
\vspace{3cm}
\epsfig{file=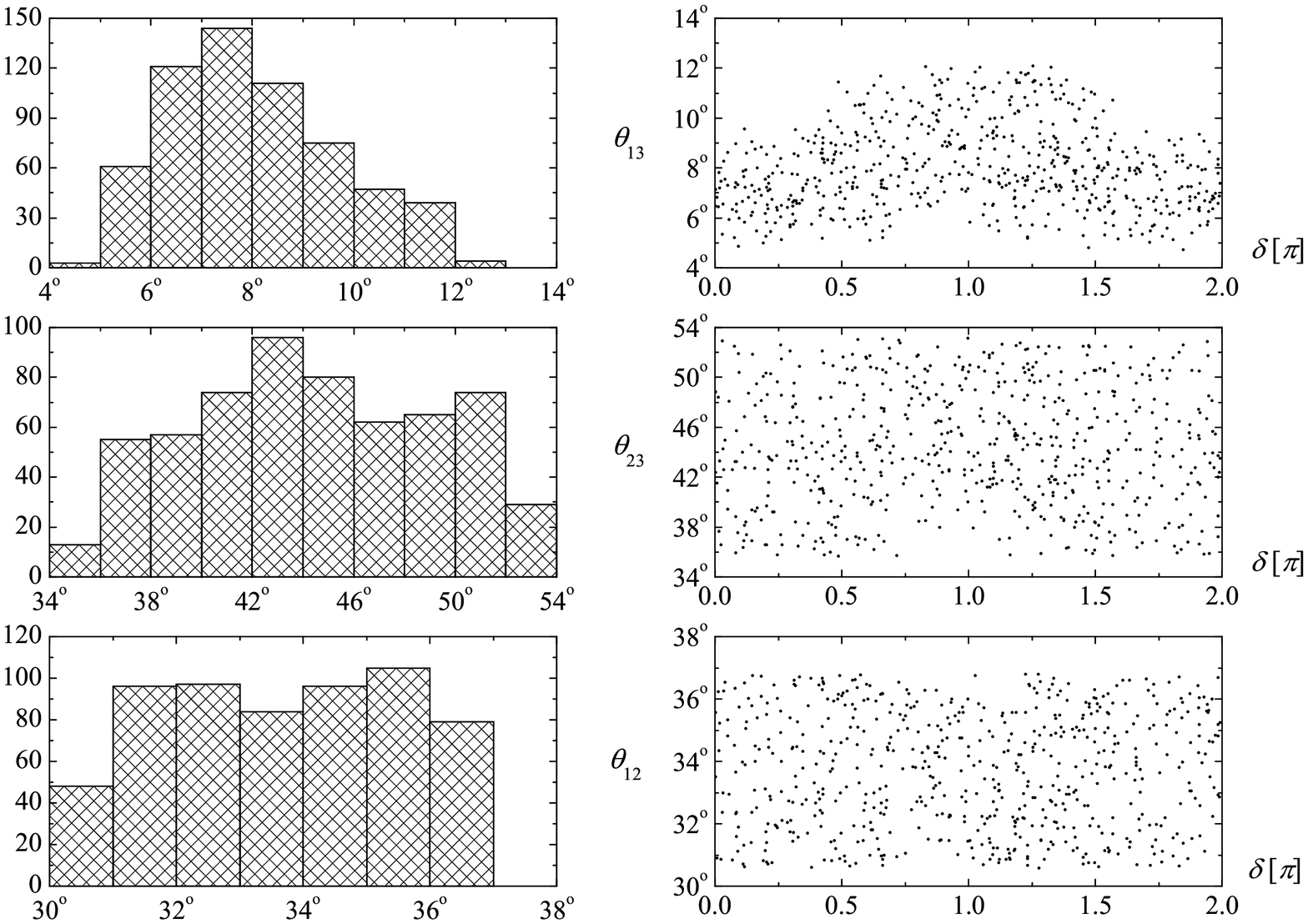,bbllx=3.1cm,bblly=12cm,bburx=8.1cm,bbury=17cm,%
width=3.5cm,height=3.5cm,angle=0,clip=0}\vspace{7cm}
\caption{Pattern $\bf A^{}_1$ of $M^{}_\nu$: allowed ranges of
flavor mixing angles $(\theta^{}_{12}, \theta^{}_{23},
\theta^{}_{13})$ versus the Dirac CP-violating phase $\delta$ at the
$3\sigma$ level, where the probability distribution of three angles
are shown in the left panel.}
\end{figure}

\newpage

\begin{figure}[b]
\vspace{3cm}
\epsfig{file=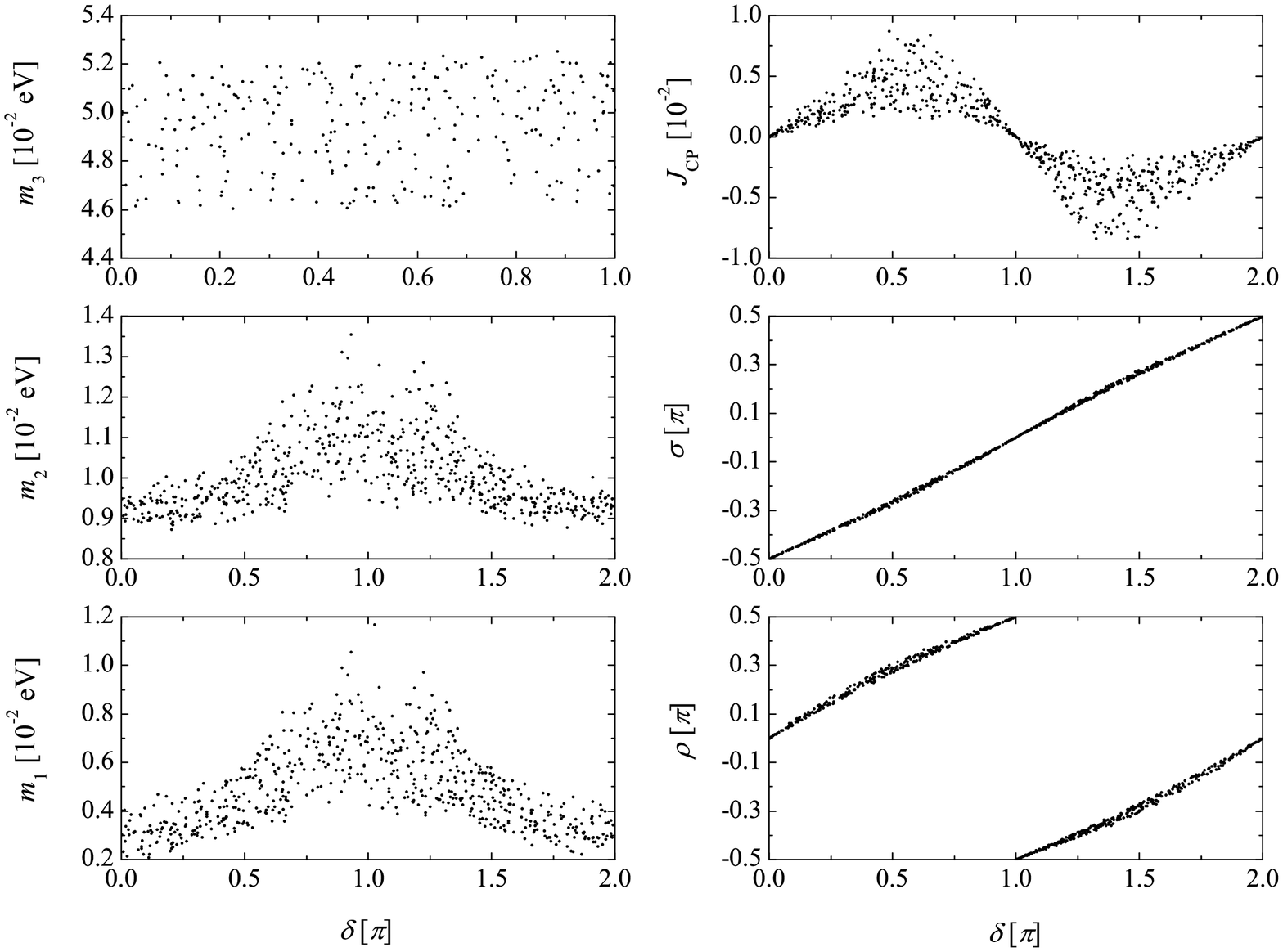,bbllx=2.2cm,bblly=12cm,bburx=7.2cm,bbury=17cm,%
width=3.5cm,height=3.5cm,angle=0,clip=0}\vspace{7cm}
\caption{Pattern $\bf A^{}_1$ of $M^{}_\nu$: allowed ranges of the
neutrino masses $(m^{}_1, m^{}_2, m^{}_3)$, the Jarlskog invariant
$J^{}_{\rm CP}$ and the Majorana CP-violating phases $(\rho,
\sigma)$ versus the Dirac CP-violating phase $\delta$ at the
$3\sigma$ level.}
\end{figure}

\newpage

\begin{figure}[b]
\vspace{3cm}
\epsfig{file=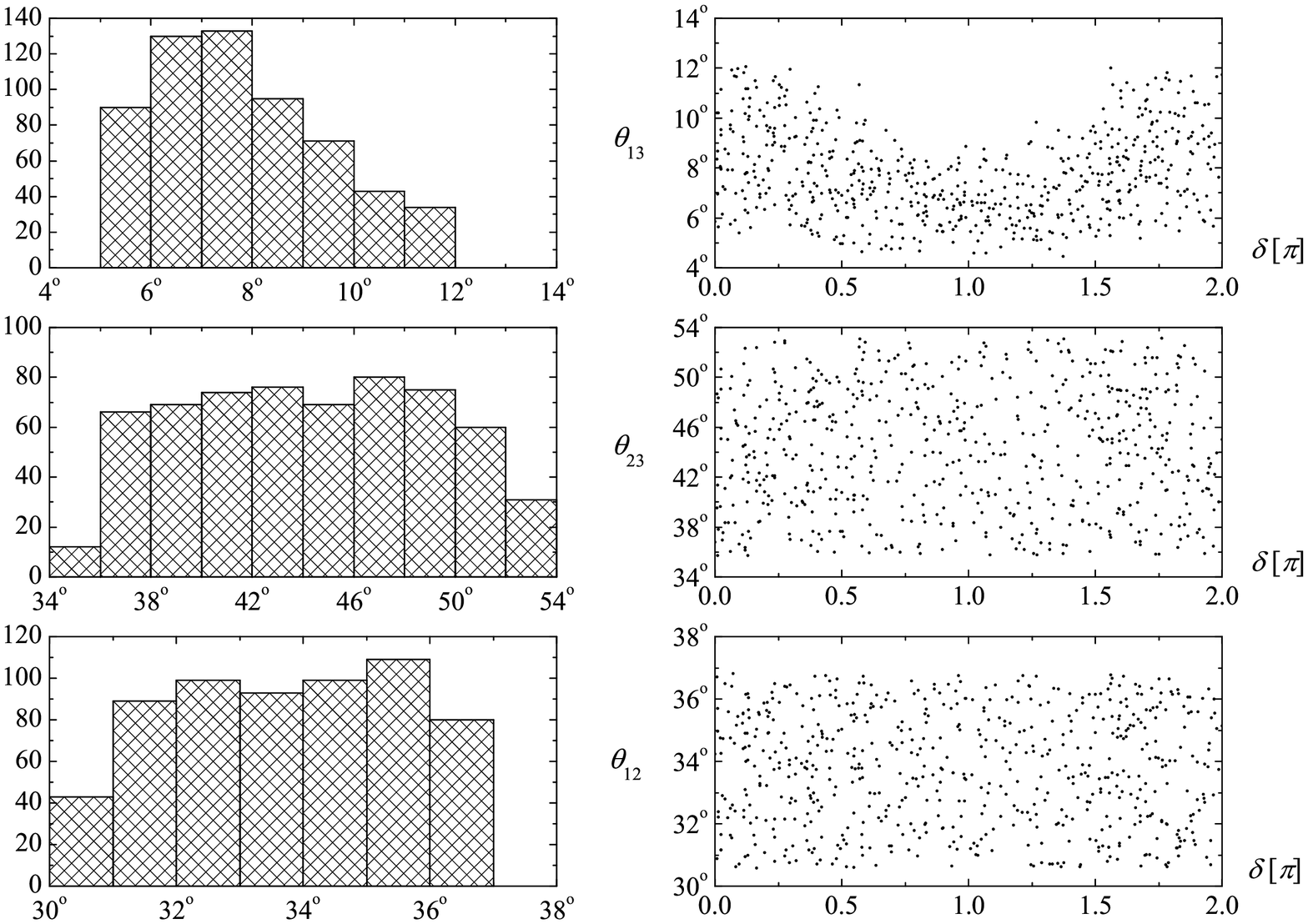,bbllx=3.1cm,bblly=12cm,bburx=8.1cm,bbury=17cm,%
width=3.5cm,height=3.5cm,angle=0,clip=0}\vspace{7cm}
\caption{Pattern $\bf A^{}_2$ of $M^{}_\nu$: allowed ranges of
flavor mixing angles $(\theta^{}_{12}, \theta^{}_{23},
\theta^{}_{13})$ versus the Dirac CP-violating phase $\delta$ at the
$3\sigma$ level, where the probability distribution of three angles
are shown in the left panel.}
\end{figure}

\newpage

\begin{figure}[b]
\vspace{3cm}
\epsfig{file=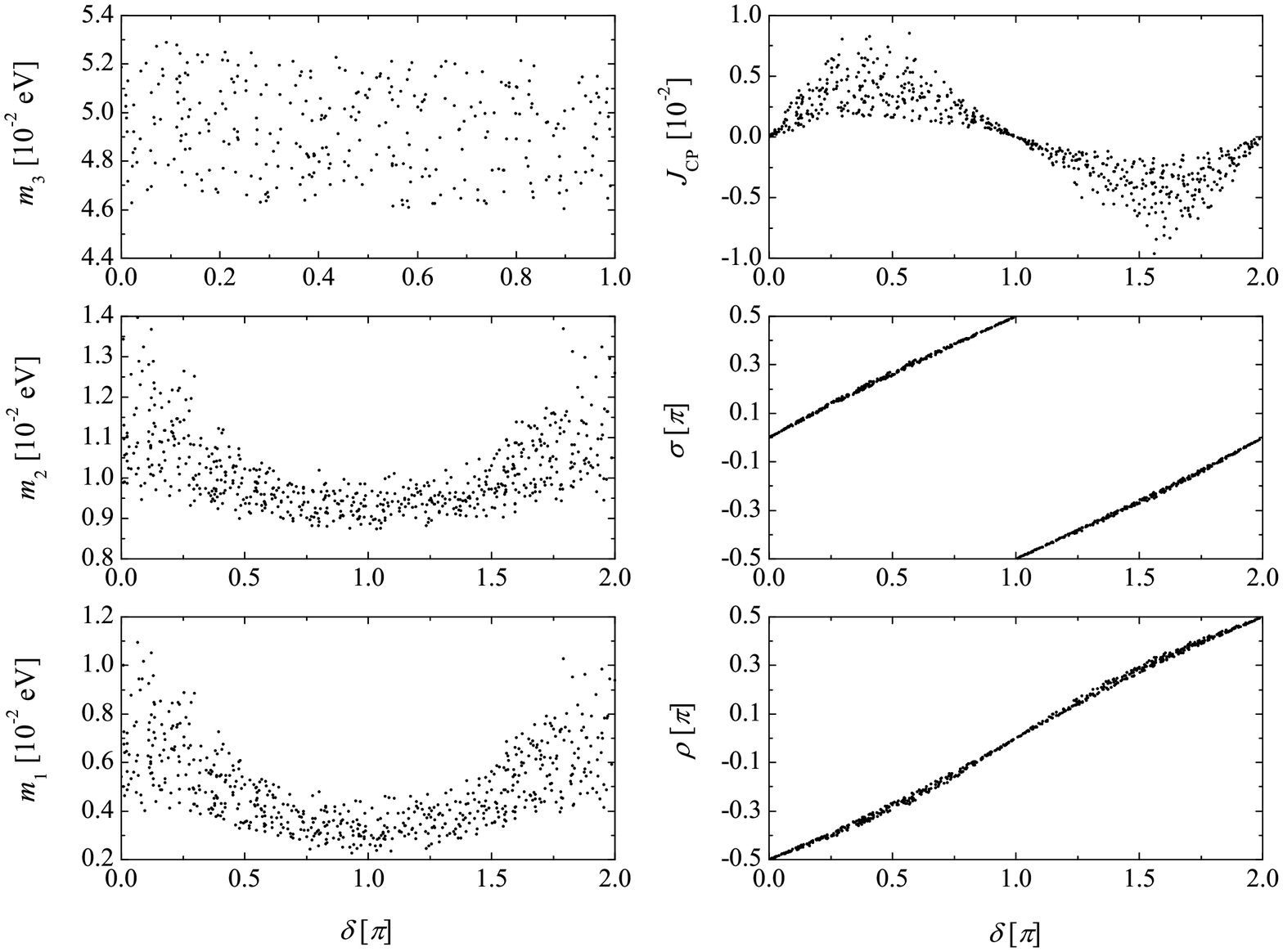,bbllx=2.2cm,bblly=12cm,bburx=7.2cm,bbury=17cm,%
width=3.5cm,height=3.5cm,angle=0,clip=0}\vspace{7cm}
\caption{Pattern $\bf A^{}_2$ of $M^{}_\nu$: allowed ranges of the
neutrino masses $(m^{}_1, m^{}_2, m^{}_3)$, the Jarlskog invariant
$J^{}_{\rm CP}$ and the Majorana CP-violating phases $(\rho,
\sigma)$ versus the Dirac CP-violating phase $\delta$ at the
$3\sigma$ level.}
\end{figure}

\newpage

\begin{figure}[b]
\vspace{3cm}
\epsfig{file=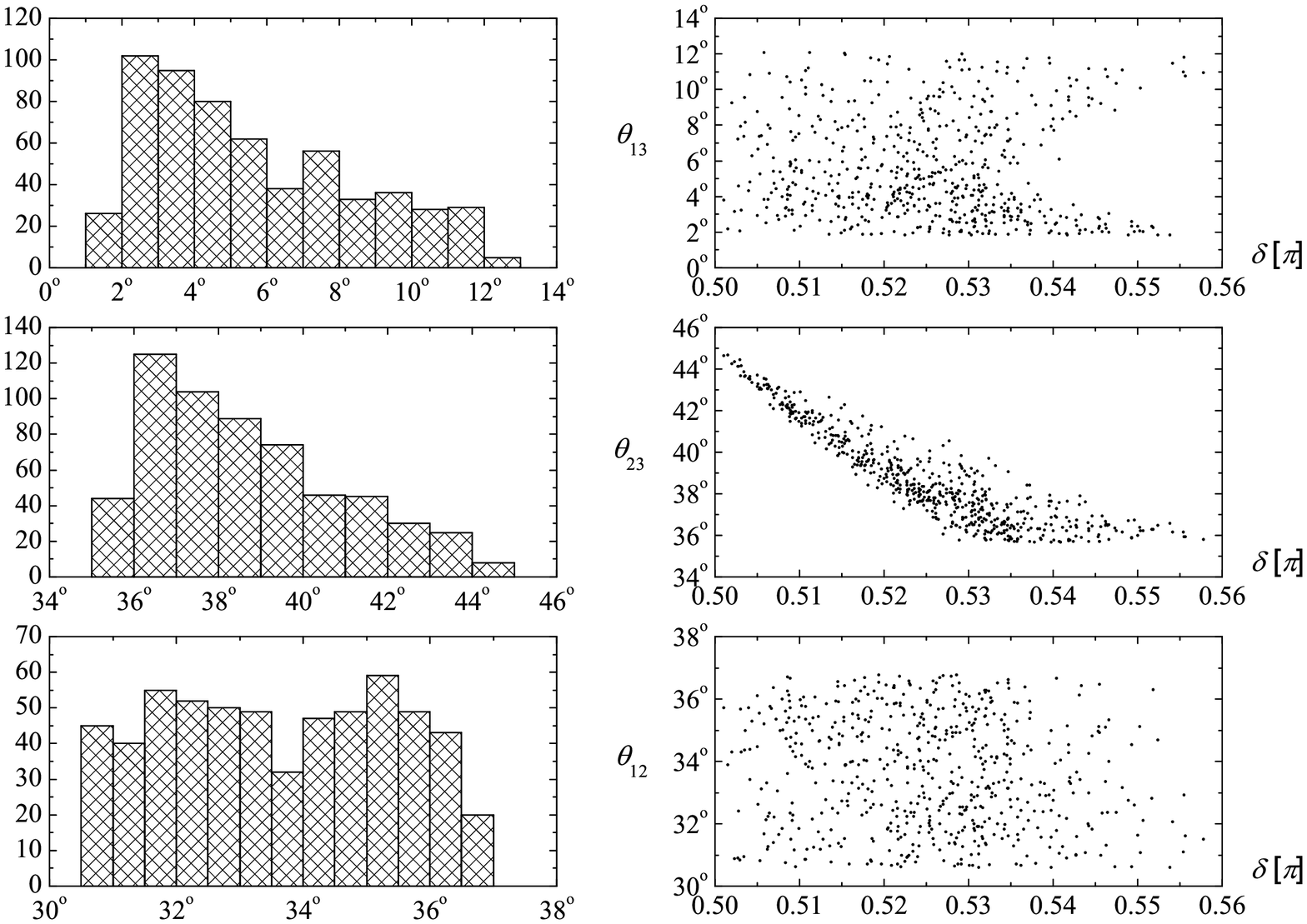,bbllx=3.1cm,bblly=12cm,bburx=8.1cm,bbury=17cm,%
width=3.5cm,height=3.5cm,angle=0,clip=0}\vspace{7cm}
\caption{Pattern $\bf B^{}_1$ of $M^{}_\nu$: allowed ranges of
flavor mixing angles $(\theta^{}_{12}, \theta^{}_{23},
\theta^{}_{13})$ versus the Dirac CP-violating phase $\delta$ at the
$3\sigma$ level, where the probability distribution of three angles
are shown in the left panel.}
\end{figure}

\newpage

\begin{figure}[b]
\vspace{3cm}
\epsfig{file=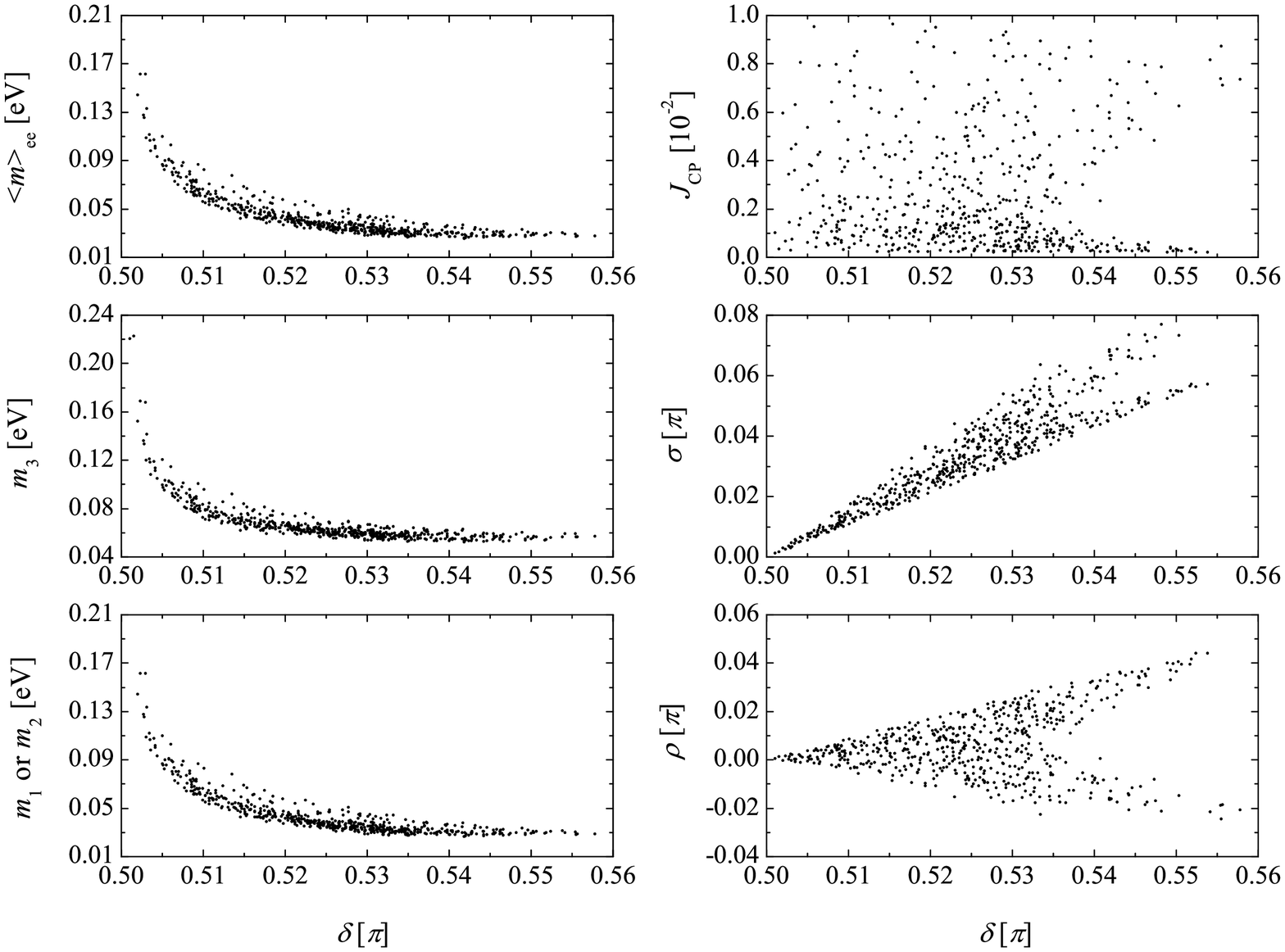,bbllx=2.2cm,bblly=12cm,bburx=7.2cm,bbury=17cm,%
width=3.5cm,height=3.5cm,angle=0,clip=0}\vspace{7cm}
\caption{Pattern $\bf B^{}_1$ of $M^{}_\nu$: allowed ranges of the
neutrino masses $(m^{}_1, m^{}_2, m^{}_3)$, the Jarlskog invariant
$J^{}_{\rm CP}$ and the Majorana CP-violating phases $(\rho,
\sigma)$ versus the Dirac CP-violating phase $\delta$ at the
$3\sigma$ level.}
\end{figure}
\newpage

\begin{figure}[b]
\vspace{3cm}
\epsfig{file=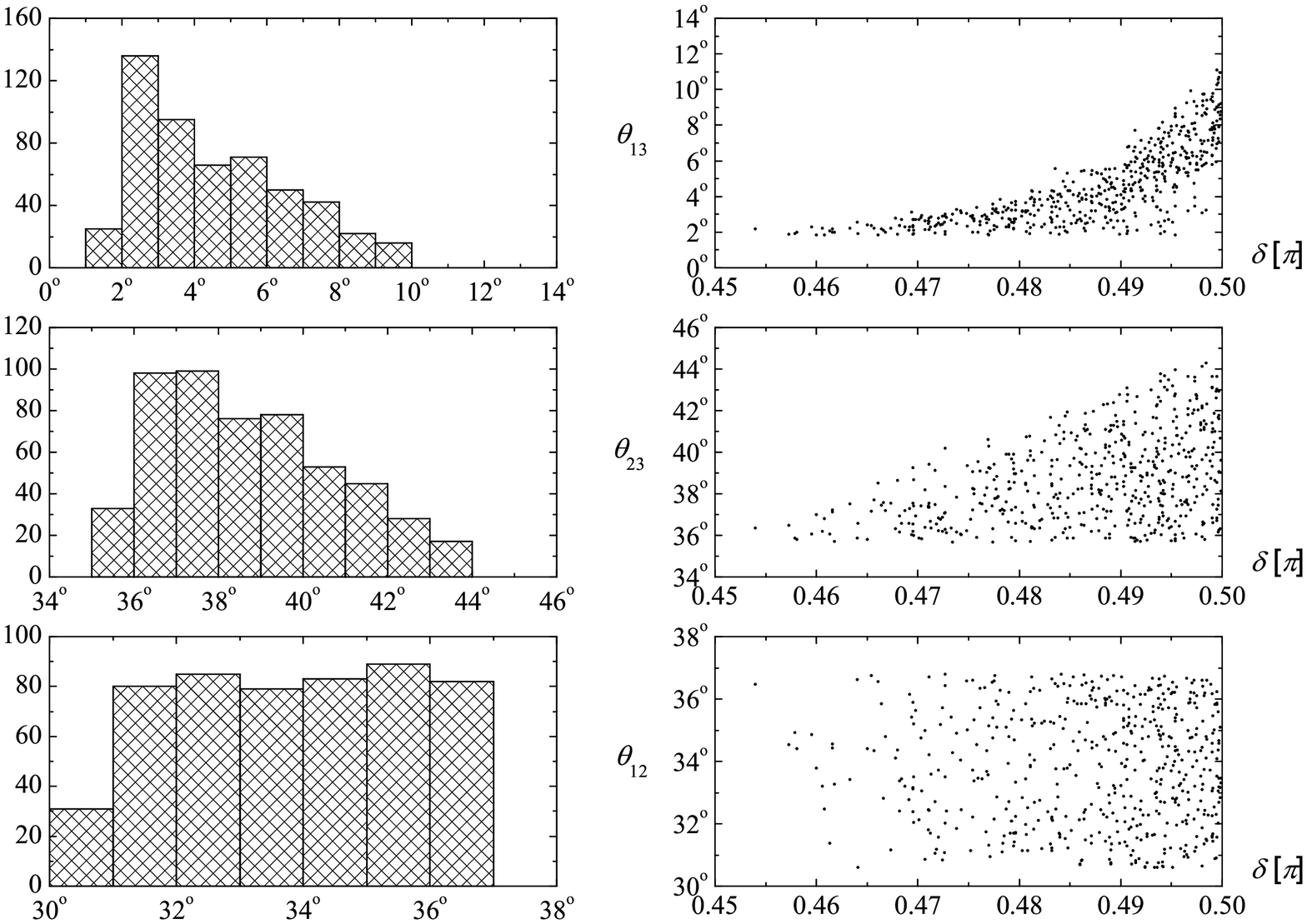,bbllx=3.1cm,bblly=12cm,bburx=8.1cm,bbury=17cm,%
width=3.5cm,height=3.5cm,angle=0,clip=0}\vspace{7cm}
\caption{Pattern $\bf B^{}_2$ of $M^{}_\nu$: allowed ranges of
flavor mixing angles $(\theta^{}_{12}, \theta^{}_{23},
\theta^{}_{13})$ versus the Dirac CP-violating phase $\delta$ at the
$3\sigma$ level, where the probability distribution of three angles
are shown in the left panel.}
\end{figure}

\newpage

\begin{figure}[b]
\vspace{3cm}
\epsfig{file=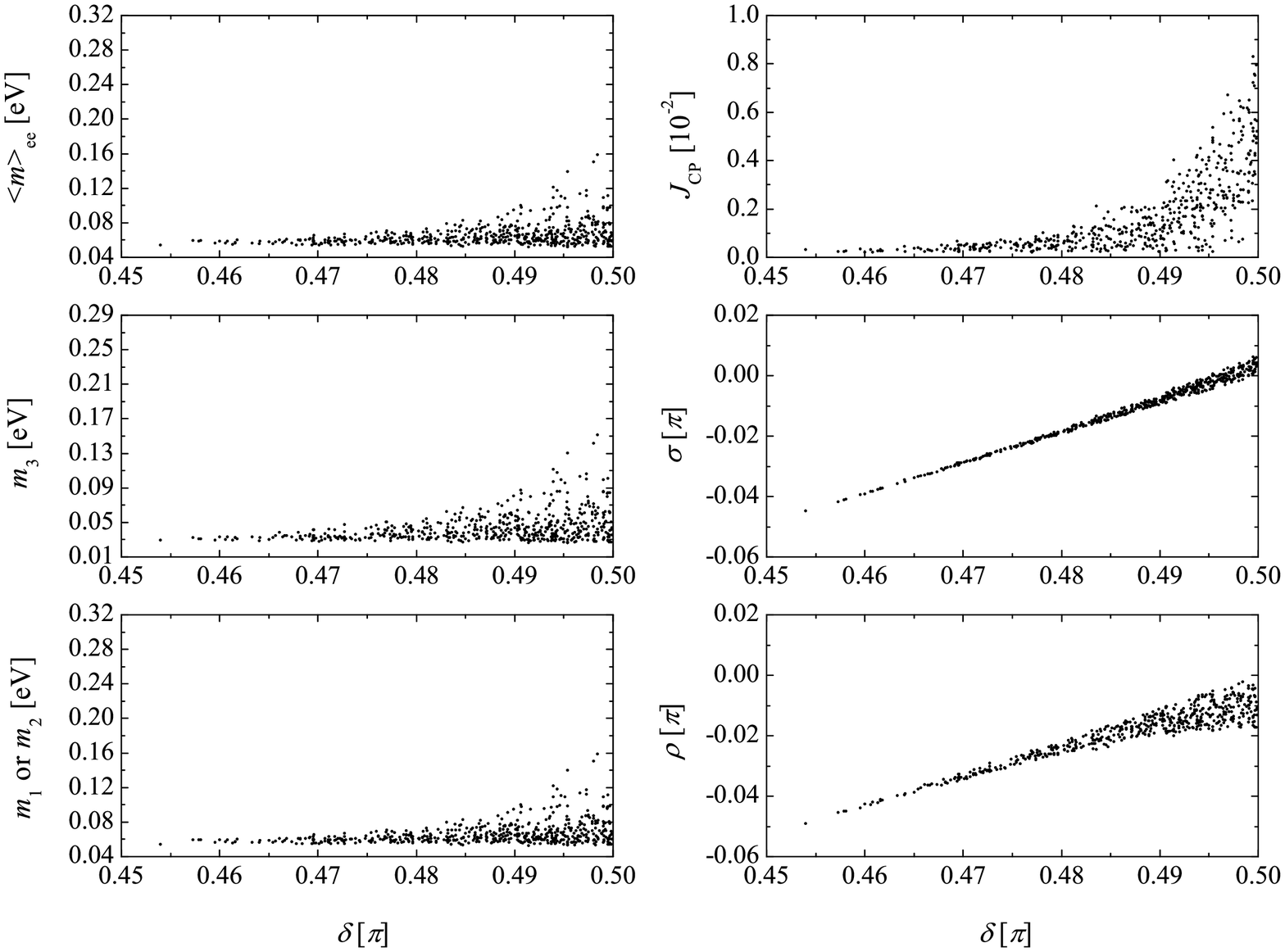,bbllx=2.2cm,bblly=12cm,bburx=7.2cm,bbury=17cm,%
width=3.5cm,height=3.5cm,angle=0,clip=0}\vspace{7cm}
\caption{Pattern $\bf B^{}_2$ of $M^{}_\nu$: allowed ranges of the
neutrino masses $(m^{}_1, m^{}_2, m^{}_3)$, the Jarlskog invariant
$J^{}_{\rm CP}$ and the Majorana CP-violating phases $(\rho,
\sigma)$ versus the Dirac CP-violating phase $\delta$ at the
$3\sigma$ level.}
\end{figure}

\newpage

\begin{figure}[b]
\vspace{3cm}
\epsfig{file=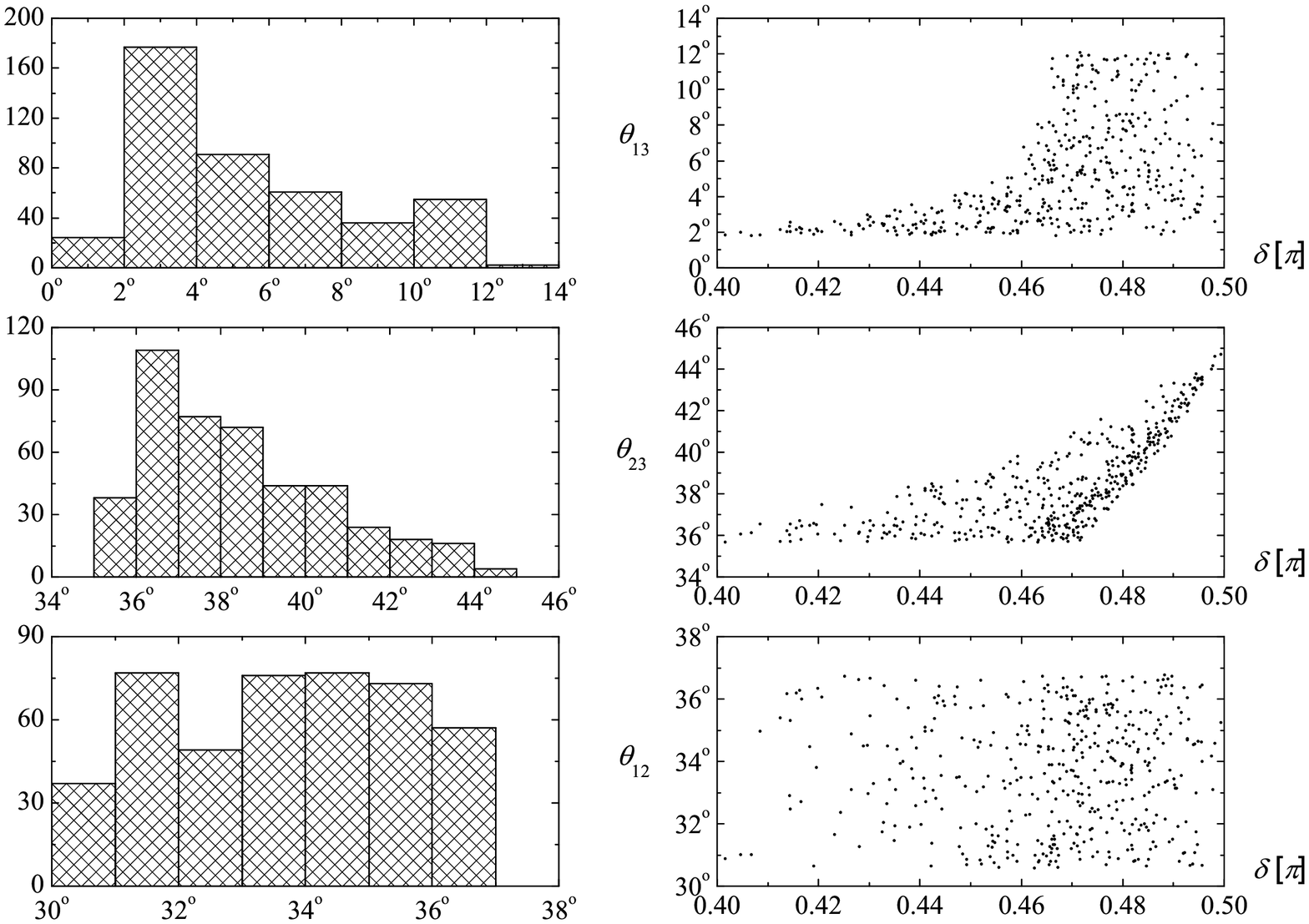,bbllx=3.1cm,bblly=12cm,bburx=8.1cm,bbury=17cm,%
width=3.5cm,height=3.5cm,angle=0,clip=0}\vspace{7cm}
\caption{Pattern $\bf B^{}_3$ of $M^{}_\nu$: allowed ranges of
flavor mixing angles $(\theta^{}_{12}, \theta^{}_{23},
\theta^{}_{13})$ versus the Dirac CP-violating phase $\delta$ at the
$3\sigma$ level, where the probability distribution of three angles
are shown in the left panel.}
\end{figure}

\newpage

\begin{figure}[b]
\vspace{3cm}
\epsfig{file=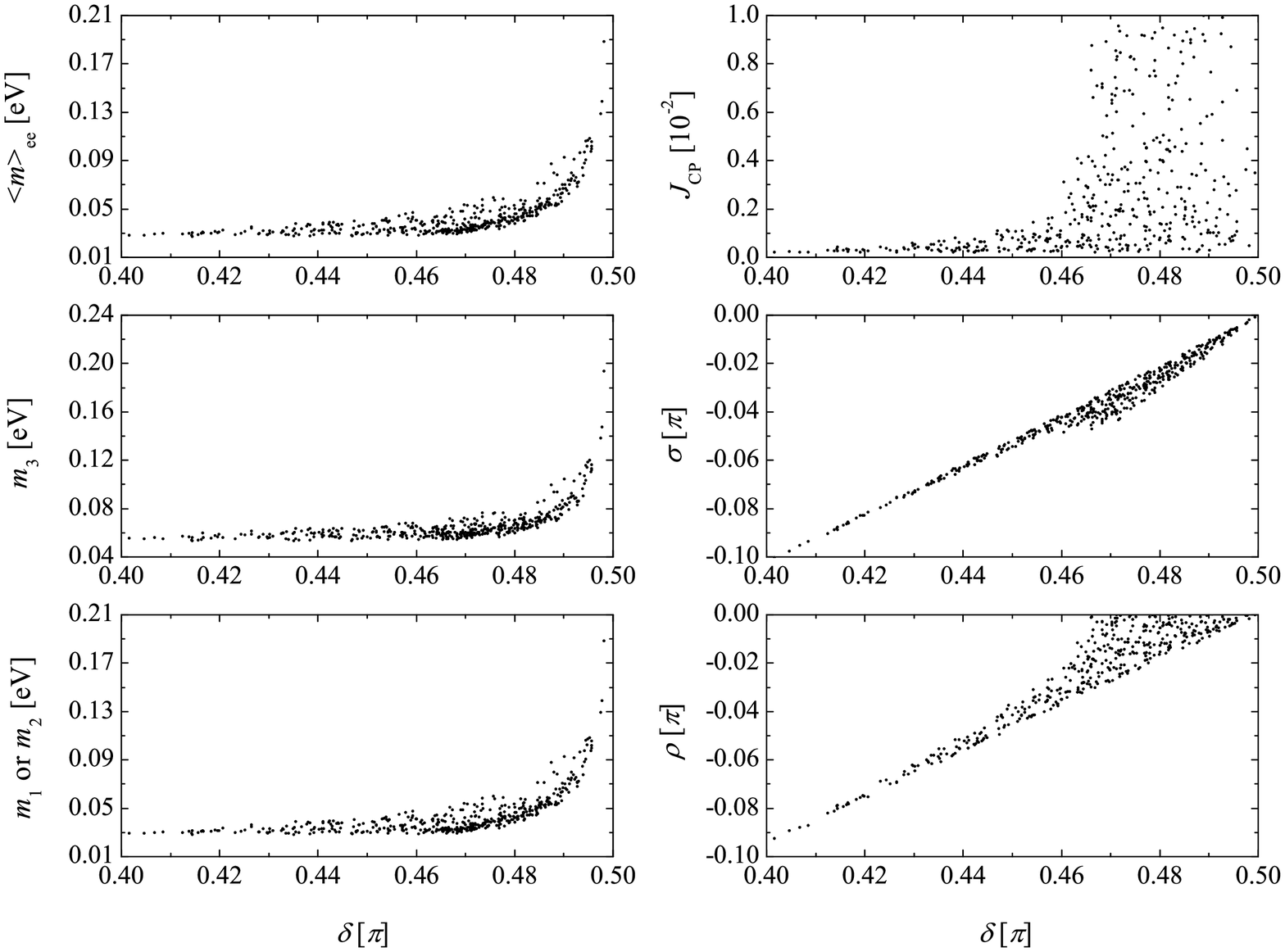,bbllx=2.2cm,bblly=12cm,bburx=7.2cm,bbury=17cm,%
width=3.5cm,height=3.5cm,angle=0,clip=0}\vspace{7cm}
\caption{Pattern $\bf B^{}_3$ of $M^{}_\nu$: allowed ranges of the
neutrino masses $(m^{}_1, m^{}_2, m^{}_3)$, the Jarlskog invariant
$J^{}_{\rm CP}$ and the Majorana CP-violating phases $(\rho,
\sigma)$ versus the Dirac CP-violating phase $\delta$ at the
$3\sigma$ level.}
\end{figure}

\newpage

\begin{figure}[b]
\vspace{3cm}
\epsfig{file=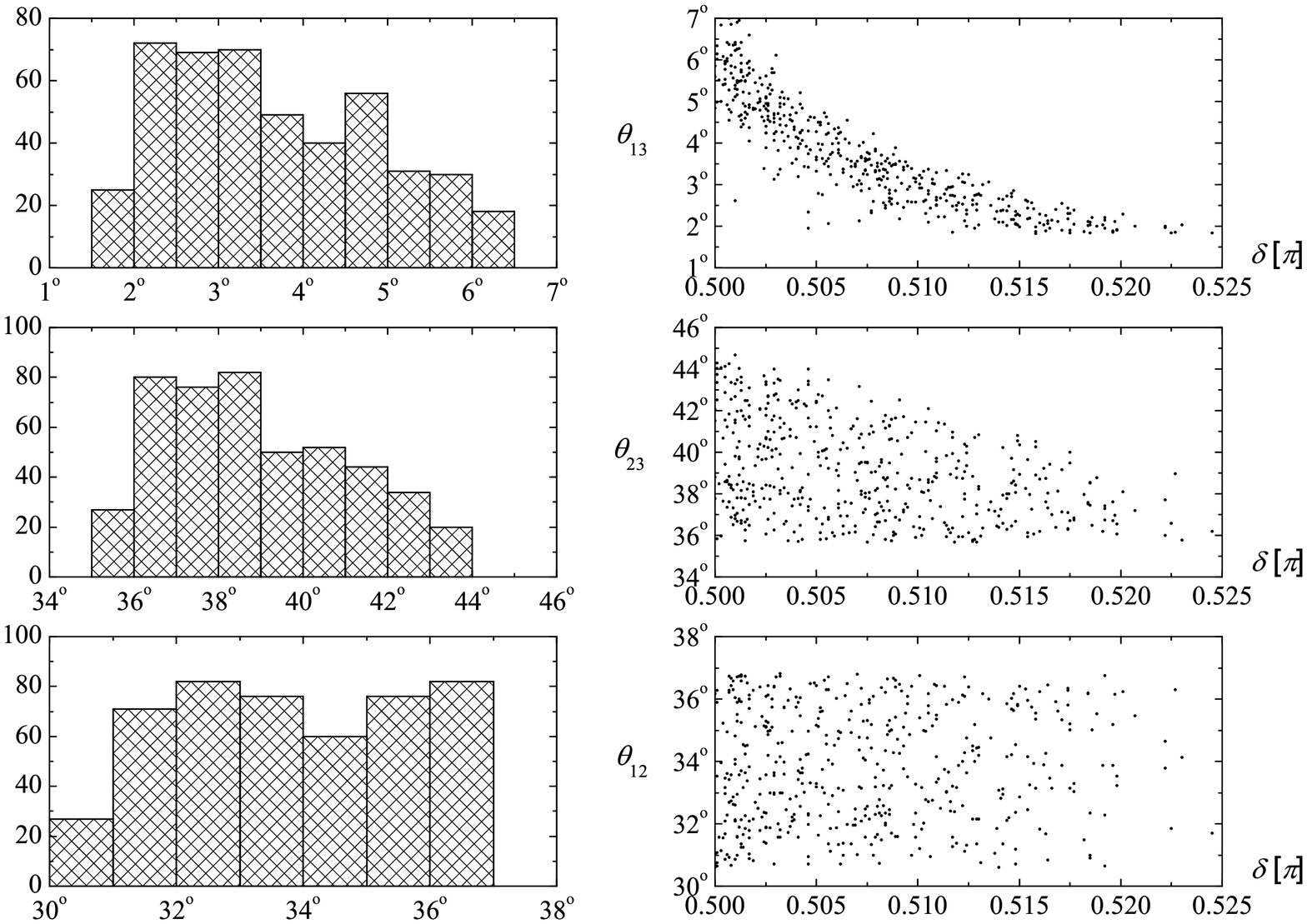,bbllx=3.1cm,bblly=12cm,bburx=8.1cm,bbury=17cm,%
width=3.5cm,height=3.5cm,angle=0,clip=0}\vspace{7cm}
\caption{Pattern $\bf B^{}_4$ of $M^{}_\nu$: allowed ranges of
flavor mixing angles $(\theta^{}_{12}, \theta^{}_{23},
\theta^{}_{13})$ versus the Dirac CP-violating phase $\delta$ at the
$3\sigma$ level, where the probability distribution of three angles
are shown in the left panel.}
\end{figure}

\newpage

\begin{figure}[b]
\vspace{3cm}
\epsfig{file=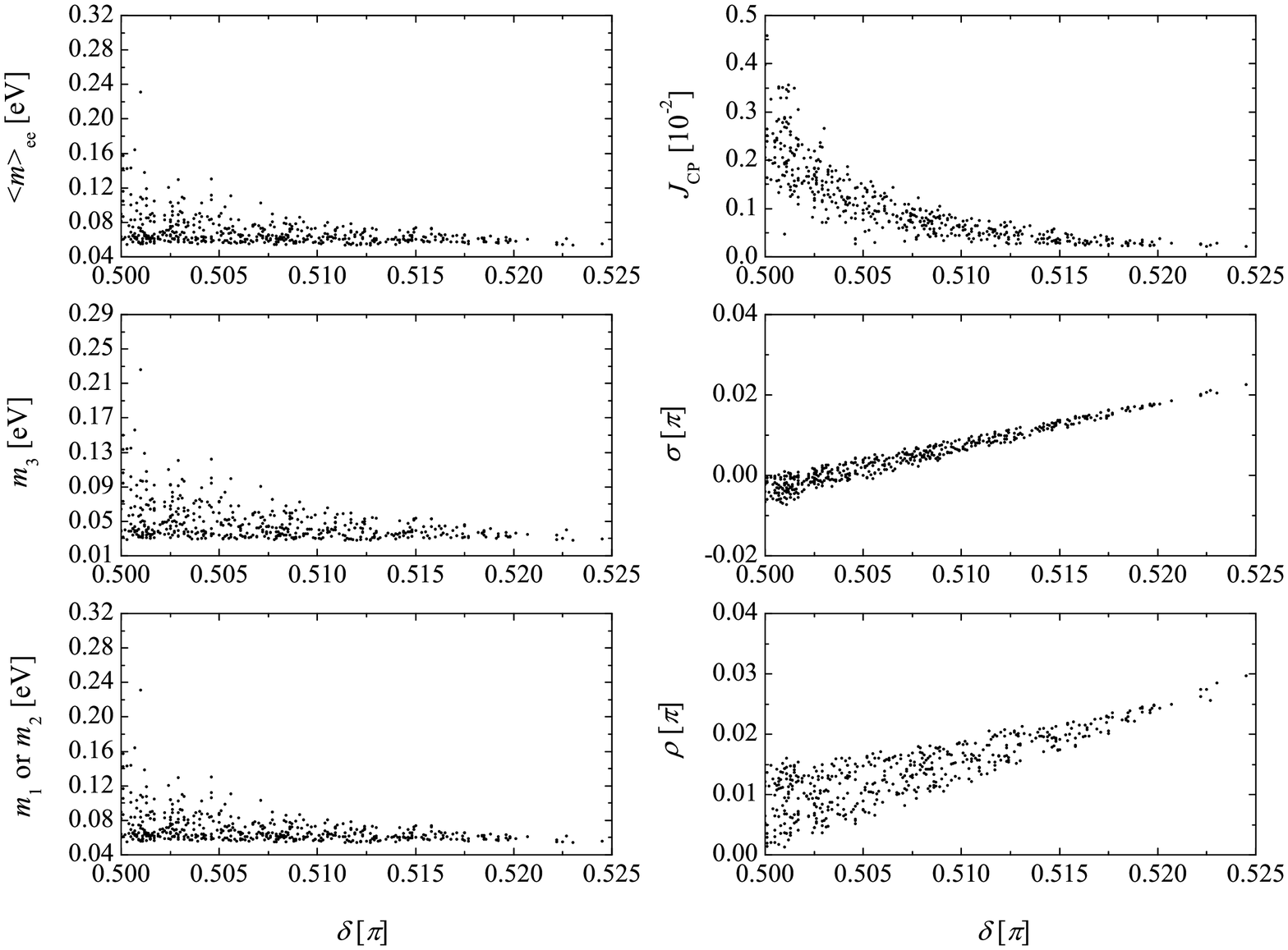,bbllx=2.2cm,bblly=12cm,bburx=7.2cm,bbury=17cm,%
width=3.5cm,height=3.5cm,angle=0,clip=0}\vspace{7cm}
\caption{Pattern $\bf B^{}_4$ of $M^{}_\nu$: allowed ranges of the
neutrino masses $(m^{}_1, m^{}_2, m^{}_3)$, the Jarlskog invariant
$J^{}_{\rm CP}$ and the Majorana CP-violating phases $(\rho,
\sigma)$ versus the Dirac CP-violating phase $\delta$ at the
$3\sigma$ level.}
\end{figure}

\newpage

\begin{figure}[b]
\vspace{3cm}
\epsfig{file=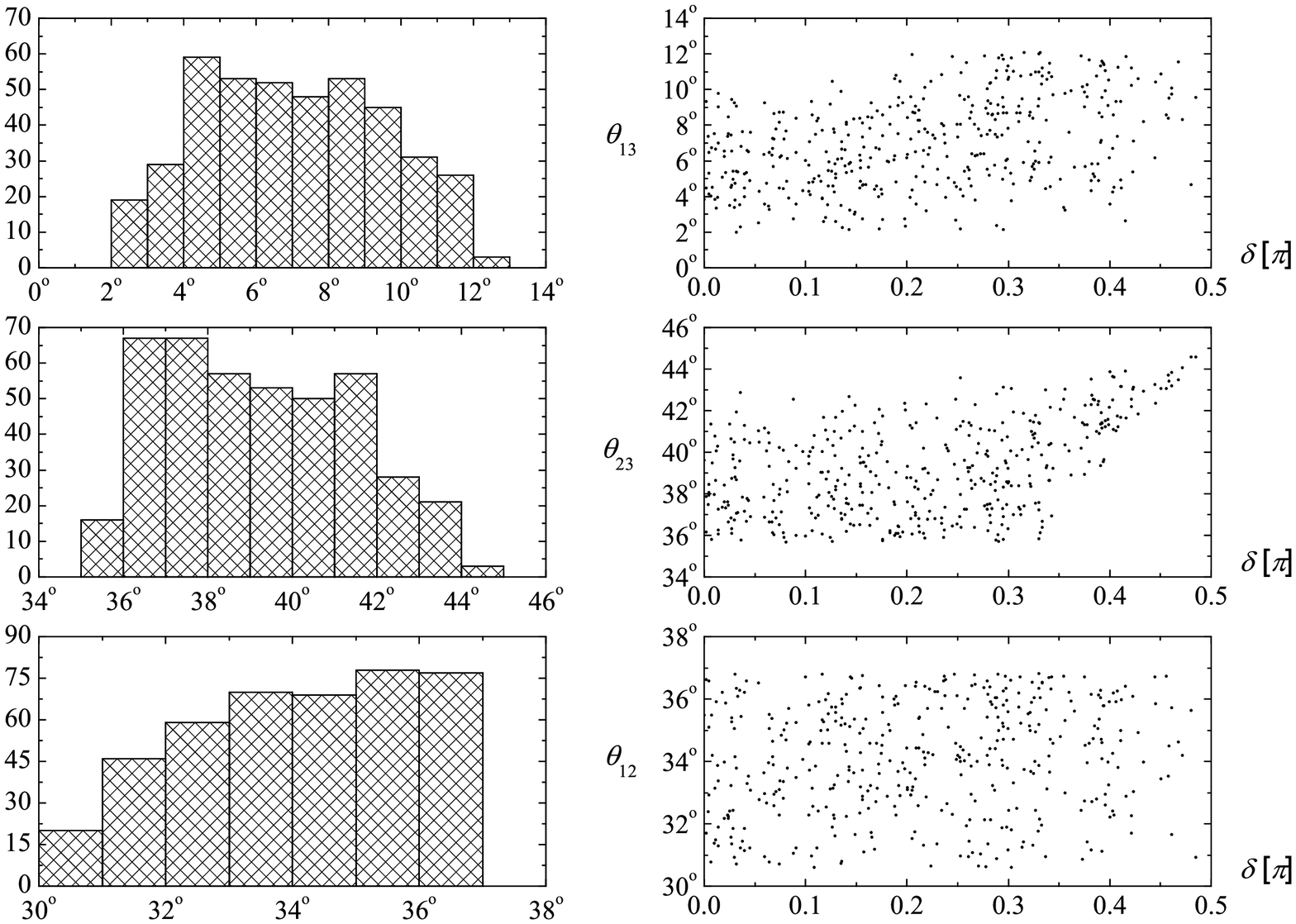,bbllx=3.1cm,bblly=12cm,bburx=8.1cm,bbury=17cm,%
width=3.5cm,height=3.5cm,angle=0,clip=0}\vspace{7cm}
\caption{Pattern $\bf C$ of $M^{}_\nu$: allowed ranges of flavor
mixing angles $(\theta^{}_{12}, \theta^{}_{23}, \theta^{}_{13})$
versus the Dirac CP-violating phase $\delta$ at the $3\sigma$ level,
where the probability distribution of three angles are shown in the
left panel.}
\end{figure}

\newpage

\begin{figure}[b]
\vspace{3cm}
\epsfig{file=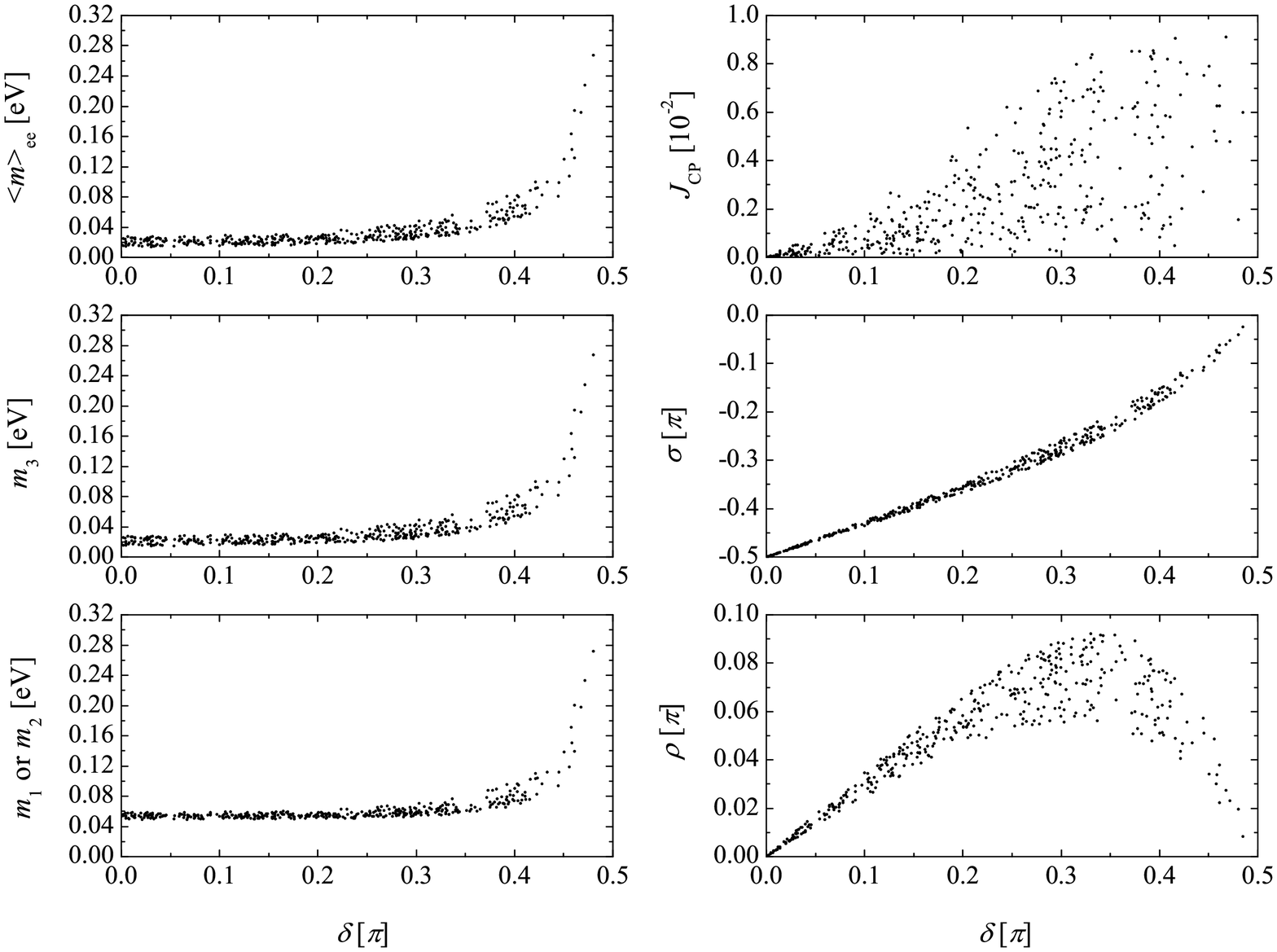,bbllx=2.2cm,bblly=12cm,bburx=7.2cm,bbury=17cm,%
width=3.5cm,height=3.5cm,angle=0,clip=0}\vspace{7cm}
\caption{Pattern $\bf C$ of $M^{}_\nu$: allowed ranges of the
neutrino masses $(m^{}_1, m^{}_2, m^{}_3)$, the Jarlskog invariant
$J^{}_{\rm CP}$ and the Majorana CP-violating phases $(\rho,
\sigma)$ versus the Dirac CP-violating phase $\delta$ at the
$3\sigma$ level.}
\end{figure}

\newpage

\begin{thebibliography}{99}

\bibitem{PDG} Particle Data Group, K. Nakamura {\it et al.},
J. Phys. G {\bf 37}, 075021 (2010).

\bibitem{T2K} T2K Collaboration, K. Abe {\it et al.},
Phys. Rev. Lett. {\bf 107}, 041801 (2011).

\bibitem{MINOS} MINOS Collaboration, P. Adamson {\it et al.},
arXiv:1108.0015.

\bibitem{DC} Double Chooz Collaboration, F. Ardellier {\it et al.},
hep-ex/0606025.

\bibitem{DYB} Daya Bay Collaboration, X. Guo {\it et al.},
hep-ex/0701029.

\bibitem{RENO} RENO Collaboration, J.K. Ahn {\it et al.},
arXiv:1003.1391.

\bibitem{X04} H. Fritzsch and Z.Z. Xing,
Prog. Part. Nucl. Phys. {\bf 45}, 1 (2000);
Z.Z. Xing, Int. J. Mod. Phys. A {\bf 19}, 1 (2004).

\bibitem{FN} C.D. Froggatt and H.B. Nielsen,
Nucl. Phys. B {\bf 147}, 277 (1979).

\bibitem{Xing04} Z.Z. Xing, hep-ph/0406049.

\bibitem{FGM} P.H. Frampton, S.L. Glashow, and D. Marfatia, Phys.
Lett. B {\bf 536}, 79 (2002).

\bibitem{xing1} Z.Z. Xing, Phys. Lett. B {\bf 530}, 159 (2002).

\bibitem{xing2} Z.Z. Xing, Phys. Lett. B {\bf 539}, 85 (2002).

\bibitem{Guo} W.L. Guo and Z.Z. Xing, Phys. Rev. D {\bf 67},
053002 (2003).

\bibitem{More} See, e.g., P.H. Frampton, M.C. Oh, and T. Yoshikawa,
Phys. Rev. D {\bf 66}, 033007 (2002); A. Kageyama, S. Kaneko, N.
Shimoyama, and M. Tanimoto, Phys. Lett. B {\bf 538}, 96 (2002); B.R.
Desai, D.P. Roy, and A.R. Vaucher, Mod. Phys. Lett. A {\bf 18}, 1355
(2003); M. Frigerio and A.Yu. Smirnov, Phys. Rev. D {\bf 67}, 013007
(2003); M. Honda, S. Kaneko, and M. Tanimoto, JHEP {\bf 0309}, 028
(2003); G. Bhattacharyya, A. Raychaudhuri, and A. Sil, Phys. Rev. D
{\bf 67}, 073004 (2003); A. Watanabe and K. Yoshioka, JHEP {\bf
0605}, 044 (2006); R. Mohanta, G. Kranti, and A.K. Giri,
hep-ph/0608292; Y. Farzan and A.Yu. Smirnov, JHEP {\bf 0701}, 059
(2007); S. Dev, S. Kumar, S. Verma, and S. Gupta, Nucl. Phys. B {\bf
784}, 103 (2007); Phys. Rev. D {\bf 76}, 013002 (2007); W.L. Guo,
Z.Z. Xing, and S. Zhou, Int. Mod. Phys. E {\bf 16}, 1 (2007); S.
Rajpoot, hep-ph/0703185; H.A. Alhendi, E.I. Lashin, A.A. Mudlei,
Phys. Rev. D {\bf 77}, 013009 (2008); E.I. Lashin and N. Chamoun,
Phys. Rev. D {\bf 78}, 073002 (2008); A. Dighe and N. Sahu,
arXiv:0812.0695; S. Goswami and A. Watanabe, Phys. Rev. D {\bf 79},
033004 (2009); S. Choubey, W. Rodejohann, and P. Roy, Nucl. Phys. B
{\bf 808}, 272 (2009); S. Dev, S. Kumar, and S. Verma, Phys. Rev. D
{\bf 79}, 033011 (2009); G. Ahuja, M. Gupta, M. Randhawa, and R.
Verma, Phys. Rev. D {\bf 79}, 093006 (2009); S. Goswami, S. Khan,
and W. Rodejohann, Phys. Lett. B {\bf 680}, 255 (2009); E.I. Lashin
and N. Chamoun, Phys. Rev. D {\bf 80}, 093004 (2009); S. Dev, S.
Verma, and S. Gupta, Phys. Lett. B {\bf 687}, 53 (2010); S. Dev, S.
Gupta, and R.R. Gautam, Phys. Rev. D {\bf 82}, 073015 (2010); W.
Grimus and P.O. Ludl, Phys. Lett. B {\bf 700}, 356 (2011).

\bibitem{Xing05} See, e.g., Z.Z. Xing, Phys. Rev. D {\bf 68}, 053002
(2003); Phys. Rev. D {\bf 69}, 013006 (2004); A. Merle and W.
Rodejohann, Phys. Rev. D {\bf 73}, 073012 (2006); Y. BenTov and A.
Zee, arXiv:1103.2616; E.I. Lashin and N. Chamoun, arXiv:1108.4010.

\bibitem{Fogli} G.L. Fogli {\it et al.}, arXiv:1106.6028.

\bibitem{Schwetz} T. Schwetz, M. Tortola, and J.W.F. Valle,
arXiv:1108.1376.

\bibitem{Weinberg} S. Weinberg, Phys. Rev. Lett. {\bf 43}, 1566
(1978).

\bibitem{SS1} P. Minkowski, Phys. Lett. B {\bf 67}, 421 (1977);
T. Yanagida, in {\it Proceedings of the Workshop on Unified Theory
and the Baryon Number of the Universe}, edited by O. Sawada and A.
Sugamoto (KEK, Tsukuba, 1979); M. Gell-Mann, P. Ramond, and R.
Slansky, in {\it Supergravity}, edited by P. van Nieuwenhuizen and
D. Freedman (North Holland, Amsterdam, 1979); S.L. Glashow, in {\it
Quarks and Leptons}, edited by M. L$\acute{\rm e}$vy {\it et al.}
(Plenum, New York, 1980); R.N. Mohapatra and G. Senjanovic, Phys.
Rev. Lett. {\bf 44}, 912 (1980).

\bibitem{SS2} W. Konetschny and W. Kummer, Phys. Lett. B {\bf 70}, 433
(1977); J. Schechter and J.W.F. Valle, Phys. Rev. D {\bf 22}, 2227
(1980); T.P. Cheng and L.F. Li, Phys. Rev. D {\bf 22}, 2860 (1980);
M. Magg and C. Wetterich, Phys. Lett. B {\bf 94}, 61 (1980); G.
Lazarides, Q. Shafi, and C. Wetterich, Nucl. Phys. B {\bf 181}, 287
(1981); R.N. Mohapatra and G. Senjanovic, Phys. Rev. D {\bf 23}, 165
(1981).

\bibitem{SS3} R. Foot, H. Lew, X.G. He, and G.C. Joshi,
Z. Phys. C {\bf 44}, 441 (1989).

\bibitem{RGE} See, e.g., P.H. Chankowski and Z. Pluciennik,
Phys. Lett. B {\bf 316}, 312 (1993); K.S. Babu, C.N. Leung, and J.
Pantaleone, Phys. Lett. B {\bf 319}, 191 (1993).

\bibitem{Mei} J.W. Mei and Z.Z. Xing, Phys. Rev. D {\bf 69}, 073003
(2004).

\bibitem{2Beta} G.L. Fogli {\it et al.}, Phys. Rev. D {\bf 78},
033010 (2008); S.M. Bilenky, Phys. Part. Nucl. {\bf 41}, 690 (2010);
W. Rodejohann, arXiv:1106.1334.

\bibitem{WMAP} E. Komatsu {\it et al.}, Astrophys. J. Suppl.
{\bf 192}, 18 (2011).

\bibitem{Grimus1} W. Grimus, A.S. Joshipura,
L. Lavoura, and M. Tanimoto, Eur. Phys. J. C {\bf 36}, 227 (2004).

\bibitem{Grimus2} W. Grimus and L. Lavoura, J. Phys. G {\bf 31}, 693
(2005); Z.Z. Xing and S. Zhou, Phys. Lett. B {\bf 679}, 249 (2009).

\bibitem{symmetry} See, e.g., M. Frigerio, S. Kaneko, E. Ma, M. Tanimoto,
Phys. Rev. D {\bf 71}, 011901 (2005); M. Hirsch, A.S. Joshipura, S.
Kaneko and J.W.F. Valle, Phys. Rev. Lett. {\bf 99}, 151802 (2007);
S. Dev, S. Gupta, and R.R. Gautam, Phys. Lett. B {\bf 701}, 605
(2011).

\bibitem{Rode} C. Hagedorn and W. Rodejohann, JHEP {\bf 0507}, 034
(2005).
\end{thebibliography}
\end{document}